\documentclass[endfloat]{geophysics}
\usepackage{graphicx}
\usepackage{subfigure}
\usepackage{amsmath,bm}
\usepackage{refcount}

\begin{document}

\title{Automatic interpretative image-focusing analysis}

\renewcommand{\thefootnote}{\fnsymbol{footnote}} 

\author{Joseph Jennings\footnotemark[2]\footnotemark[1], 
Robert Clapp\footnotemark[2], 
Mauricio Araya-Polo\footnotemark[3] 
and Biondo Biondi\footnotemark[2]}

\footnotetext[2]{Stanford University, Department of Geophysics, Stanford, CA, U.S.A}
\footnotetext[3]{Total Energies, Research and Technology US, Houston, TX, U.S.A}
\footnotetext[1]{corresponding author: joseph29@stanford.edu}

\footer{Interpretative image-focusing analysis}
\lefthead{Jennings et. al}
\righthead{Interpretative image-focusing analysis}

\maketitle

\begin{abstract}
  The focusing of a seismic image is directly linked
  to the accuracy of the velocity model. Therefore,
  a critical step in a seismic imaging workflow is to perform a
  focusing analysis on a seismic image to determine velocity errors.
  While the offset/aperture-angle axis is frequently
  used for focusing analysis, the physical (i.e., midpoint) 
  axes of seismic images tend to be ignored
  as focusing analysis of geological structures is highly interpretative
  and difficult to automate. We have developed an automatic data-driven
  approach using convolutional neural networks 
  to automate image-focusing analysis. Using focused and unfocused geological faults,
  we show that our method can make use of both spatial and offset/angle focusing
  information to robustly estimate velocity errors within seismic 
  images. We demonstrate that our method correctly estimates velocity errors
  from a 2D Gulf of Mexico limited-aperture image where a traditional 
  semblance-based approach fails.
  We also show that our method has the added benefit of improving 
  the interpretation of faults within the image.
\end{abstract}

\section*{Introduction}

The focusing and unfocusing of seismic images is in large part 
determined by the accuracy of the migration velocity model. When the migration
velocity is incorrect, within their physical axes ($x$-midpoint, $y$-midpoint and depth),
seismic images lose reflector coherence and in the presence of geological 
faults and rugose geological features, exhibit unfocused/overly 
focused fault plane reflections and diffractions.
The angle/offset axis within prestack seismic images provides an additional axis
for analyzing seismic image focusing. When the migration velocity is correct, events within
angle/offset gathers will be flat; when it is incorrect, they will either curve upward
or downward.
The use of focusing information within the physical axes as well as the angle/offset axis
for image-focusing analysis has shown to provide high resolution and robust estimation
of migration velocities in scenarios with limited aperture-angle range such as 
sub-salt imaging as well as narrow/constant-offset acquisition geometries 
\cite[]{sava2005wave,wang20063d,wang2008advances,ma2011enhanced}.
Notwithstanding the success obtained with using all axes of a seismic image, 
typically velocity errors are assessed within seismic images
using only the residual moveout information along the angle/offset axis.
This is mainly due to the fact that analyzing focusing within the physical axes of
seismic images is a highly interpretative process that is difficult to automate.
Moreover, it tends to be more applicable in areas of with rugose and faulted
geology which occur less frequently in the subsurface.

In spite of this challenge, many successful attempts have been made to 
create a robust and automatic image-focusing analysis. Existing approaches
can be split into two main categories. The first of these categories
are diffraction-based approaches that 
analyze the focusing of diffractions separated from the reflections. These approaches
have been well-studied and have been able to improve imaging and interpretation
in a variety of geological settings such as near
salt bodies, faults and also channels 
\cite[]{de1984velocity, harlan1984signal,fomel2007poststack, montazeri2020improved}.
The majority of these
approaches only examine the focusing of a seismic image within its physical axes
and use an image entropy-based focusing measure for determining the optimal
focusing of a diffraction.

Approaches that fall into the other category make use of coherence along the different axes 
of the image.  Within images generated from common-reflection point (CRP)
scans \cite[]{audebert1996crp}, wave-equation migration scans 
\cite[]{wang20063d} or reverse-time migration delayed-imaging-time scans 
\cite[]{wang2009subsalt}, they examine either the flatness of the CRP gathers 
or the coherence of extracted horizons. 
While original approaches analyzed the coherence of horizons qualitatively 
\cite[]{negron20003d, wang20063d}, 
automatic methods were later developed to perform focusing analysis for each image point 
along automatically extracted horizons \cite[]{whiteside2011automatic}. 
These approaches have shown to provide 
robust estimates of focusing errors within images and also demonstrated 
to have the added benefit of improving the interpretation of geologic features 
such as salt bodies. 

While approaches in both of these categories have demonstrated success towards developing
an automated interpretative image-focusing analysis, their solutions are not all-encompassing.
The diffraction-based approaches rely on the detection and isolation of diffractions from 
reflections within seismic images, which can prove to be a challenging task in the presence
of noise or limited illumination \cite[]{berkovitch2009diffraction}. 
Additionally, these approaches are limited in that they
only target point scatterers within stacked seismic images. Therefore, they neglect reflector
coherence and curvature in addition to the curvature information 
available along the angle/offset axis of prestack seismic images. 
While the coherence-based approaches have become routine procedures
for subsalt-imaging workflows, they are also not without limitations. Although
they incorporate structural constraints by performing the focusing analysis along extracted
horizons, the picking of the optimal focusing parameters 
is typically performed on each image-point
separately using a maximum amplitude criterion, which can potentially lead to a spatially
inconsistent assessment of the focusing errors. Additionally, both the diffraction-based
and the coherence-based approaches can be considered from an automation perspective as 
``rule-based" approaches that rely on the specification and extraction of particular
features within focused and unfocused images in order to assess the velocity error.
While these features can be extracted and used for a wide variety of seismic images,
there are also many instances in which they do not apply (e.g., corner cases) and
therefore other features need to specified and the rule-based approaches need to be updated.

In this paper, we present a novel workflow of automatic interpretative image-focusing analysis
that uses the powerful automatic feature extraction capabilities of deep learning with
convolutional neural networks (CNNs). Recently, CNNs have been transforming the seismic
interpretation community due to their ability to robustly detect geological features
such as geological faults, channels and salt bodies within seismic images
\cite[]{waldeland2018convolutional, wu2019faultseg3d, pham2019automatic, tschannen2020extracting}.
The CNN we present and train within this study is similar to the work done in the seismic
interpretation community. While CNNs trained for seismic interpretation typically learn to 
detect the presence of geological features within seismic images, we train a CNN to learn
to detect whether a geological feature is focused within an image. Although any geological feature
could be used for training the CNN, for the purposes of this paper, we provide the CNN
with focused and unfocused faults as training images. Our image-focusing CNN takes as input
prestack image patches from a migration scan and provides for each image patch a focusing
score that is close to one for focused images and close to zero for unfocused images.
With focusing scores computed for all patches in the image, the migration scanning parameter can
then be selected for each patch based on the image that provided the largest focusing score.
While any migration scanning procedure can be used in this workflow, in this study, we use
prestack Stolt residual depth migration for generating focused and unfocused prestack
images \cite[]{sava2003prestack, biondi2010velocity}.

To demonstrate the effectiveness of our method, we train our image-focusing CNN and use it
to automate an image-focusing analysis on a 2D limited-aperture image from the Gulf of
Mexico (GOM). Comparing our approach to an approach that only analyzes the migration
scans using semblance calculated from angle-domain common image gathers, we show that in 
spite of the limited amount of angle information provided along the aperture-angle axis,
our method can make use of the additional spatial focusing information from faults
to provide a robust estimate of the migration scanning parameter. We also show that
our method, similar to \cite{whiteside2011automatic}, has the added benefit of improving
the interpretation of the faults compared to the semblance-based method.

In the first section of the paper, we introduce the 2D limited-aperture GOM field dataset 
to which we apply image-focusing analyses using both a semblance-based approach 
and our novel CNN-based approach. The image obtained from migrating these data
is composed primarily of normal listric faults and therefore 
the geological feature on which we choose to 
spend most of our attention for focusing analysis is faults. 
Following the description of the field data, we review the migration
scanning procedure of prestack Stolt residual migration and the
semblance-based approach for performing focusing analysis that only uses the 
aperture angles.
We then outline how we perform the focusing
analysis using a CNN by describing design of the CNN, the training data creation workflow and 
the training of the CNN. Finally, we apply both approaches to our 2D GOM image with faults
and compare and discuss the results.

\section*{Field data}

\begin{figure}
  \includegraphics[width=\textwidth]{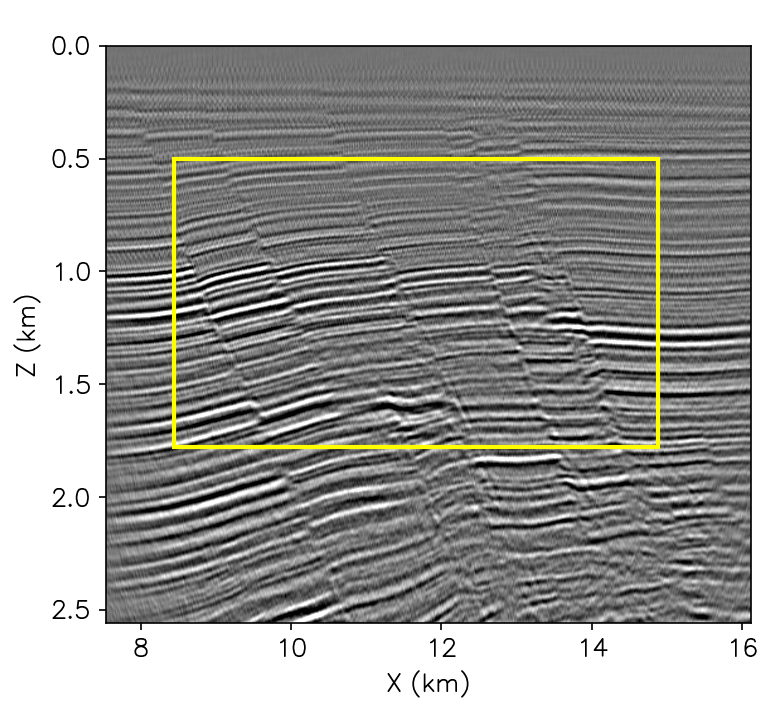}
  \caption{Unfocused depth-migrated image to be used in this study with the 
  region of interest for focusing analysis shown within the highlighted box.}
  \label{fig:imgbox}
\end{figure}

The field data used for this study were taken from a survey line acquired off of the
Texas coast of the Gulf of Mexico \cite[]{claerbout1985imaging}. The data consist of 148 shots
acquired with a streamer acquisition with a maximum offset of 3.4 km. Figure \ref{fig:imgbox}
shows the migrated image of these data. We migrated the dataset
with a 2D shot-profile wave-equation depth migration algorithm 
\cite[]{kessinger1992extended} that used a maximum frequency of 50 Hz and 41 subsurface offsets. 
During the migration, we introduced a shallow slow velocity anomaly 
in order to introduce defocusing on the normal faults present within the image. 

\begin{figure}
  \includegraphics[width=\textwidth]{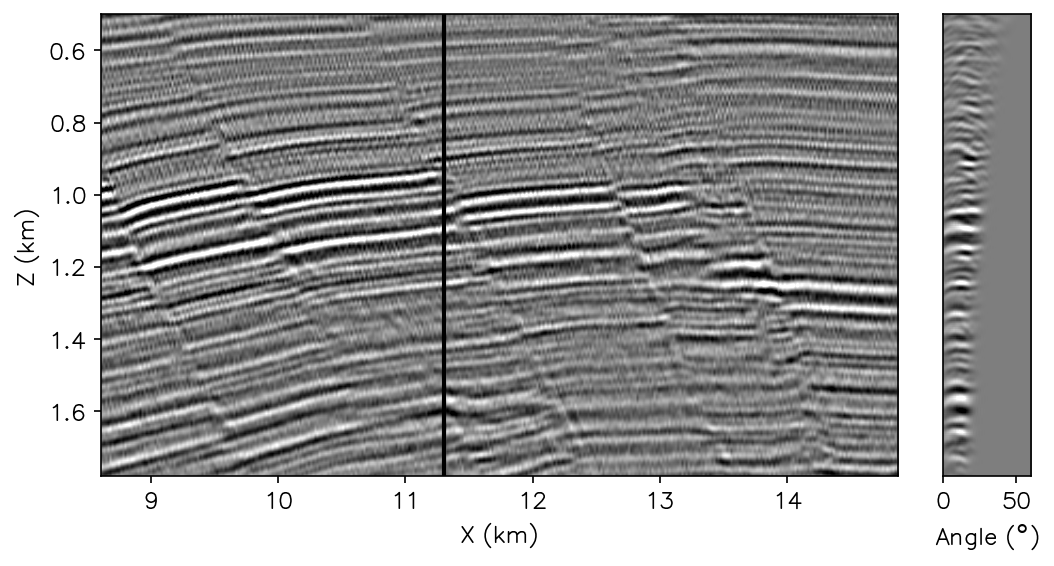}
  \caption{The region of interest shown in Figure 
  \ref{fig:imgbox} and a muted angle gather extracted at 11.3 km.}
  \label{fig:zoomimgmute}
\end{figure}

Additionally, in order to investigate the effects of a limited aperture during image 
focusing analysis, we converted the subsurface-offset-domain common-image gathers
to aperture-angle-domain common-image gathers \cite[]{sava2003angle} and
applied a mute on the angle gathers in order to simulate data that had been acquired 
with only 1.0 km offset. Figure 
\ref{fig:zoomimgmute} shows the region of interest of the image used for this study as 
well as an extracted muted angle gather. Diffracted events on the fault near $x=10$ km 
and the unfocused fault plane reflection on the central fault provide signs of 
undermigration. Also note that little to no curvature information is available along the aperture 
angle axis due to the application of the mute.

\section*{Methods}

\subsection*{Prestack Stolt residual depth migration}

We use a residual migration-based approach for performing the migration scanning procedure.
Residual migration is a seismic imaging technique that transforms an image that has 
been migrated with a particular velocity, to an image that has been migrated with 
another velocity. A major advantage of this image-to-image mapping is that as 
residual velocity errors are in general small, less accurate imaging algorithms
may be utilized \cite[]{rothman1985residual}. 
Typically, less accurate imaging algorithms are also less computationally
demanding and therefore many residual migrations can be performed at relatively little cost.
Over the past few decades, residual migration has shown to be a useful tool for imaging
and migration velocity analysis 
\cite[]{beasley1988cascaded,van1990tomographic,sava2004wave,novais2008gpr,alkhalifah2011basic}.

While any migration can be re-expressed as a residual migration algorithm, in this study 
we use the 2D prestack Stolt residual depth migration algorithm as described in 
\cite{sava2003prestack}. 
This migration algorithm performs a mapping from 
our initial migration image (migrated with initial velocity $v_0(z, x)$)
to a residually migrated image (migrated with trial velocity $v(z, x)$):
\begin{equation}
  r_0(z_0, x, h) \rightarrow r(z, x, h),
\end{equation}
where $z$ is depth, $x$ is midpoint and $h$ is subsurface offset. 
This mapping is performed in the spatial frequency domain using the following relationship:
\begin{align}
 k_z =& \frac{1}{2}\sqrt{\frac{v_0^2}{v^2}\frac{\left[k_{z_0}^2 + k_{h}^2\right]
 \left[k_{z_0}^2 + k_{x}^2\right]}{k_{z_0}^2} - (k_{x} + k_{h})^2} \notag \\
 & \quad\quad\quad+ \frac{1}{2}\sqrt{\frac{v_0^2}{v^2}\frac{\left[k_{z_0}^2 + k_{h}^2\right]
 \left[k_{z_0}^2 + k_{x}^2\right]}{k_{z_0}^2} - (k_{x} - k_{h})^2},
 \label{eqn:restolt}
\end{align}
where $k_z$ is the vertical wavenumber of the output residually migrated image, $k_{z_0}$
is the vertical wavenumber of the original input image, $k_{h}$ is the subsurface offset
wavenumber, and $k_{x}$ is the midpoint wavenumber. It is important to note that Equation 
\ref{eqn:restolt}
does not only depend on the original migration velocity, but rather on the ratio between 
the original migration velocity and the output trial velocity. Following the notation of 
\cite{biondi20063d}, we denote this parameter as $\rho$ and refer to it throughout this paper
as the migration scanning parameter. As prestack Stolt residual depth
migration operates entirely in the wavenumber domain (described by Equation \ref{eqn:restolt}),
the ratio $\rho = v_0/v$ must be a scalar parameter. 
This implies that while $v_0$ and $v$ may both
be spatially-variant, $v$ must be a scaled version of $v_0$. In this sense, performing prestack 
Stolt residual migration for a range of $\rho$ values is very similar to performing wave equation
migration/CRP scans for focusing migration velocity analysis.

\begin{figure}
  \subfigure[]{\includegraphics[width=0.5\textwidth]{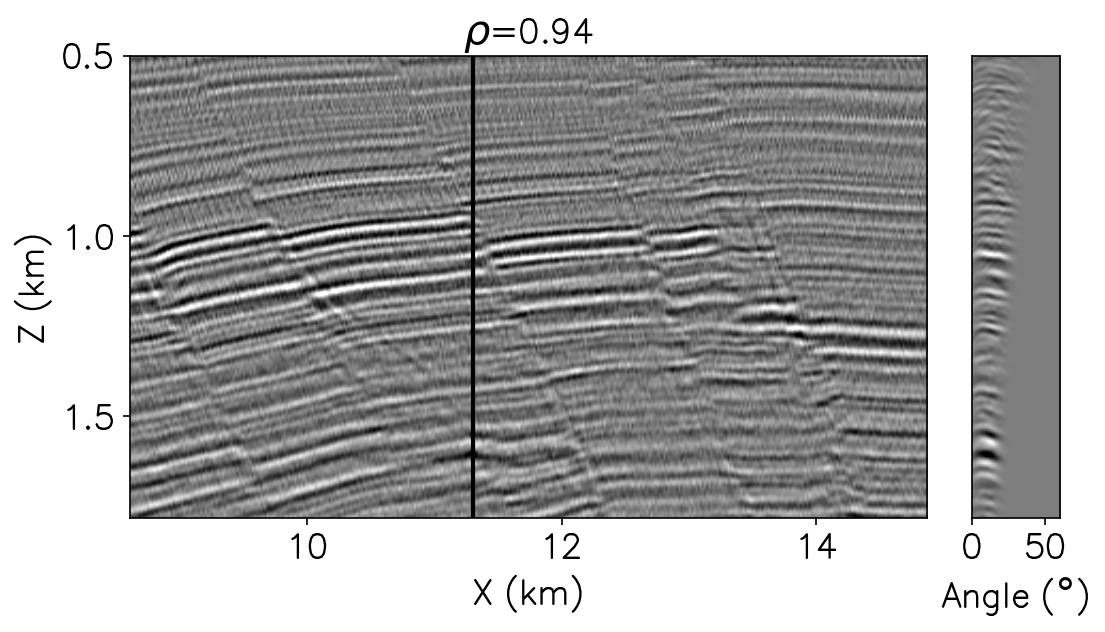}}
  \subfigure[]{\includegraphics[width=0.5\textwidth]{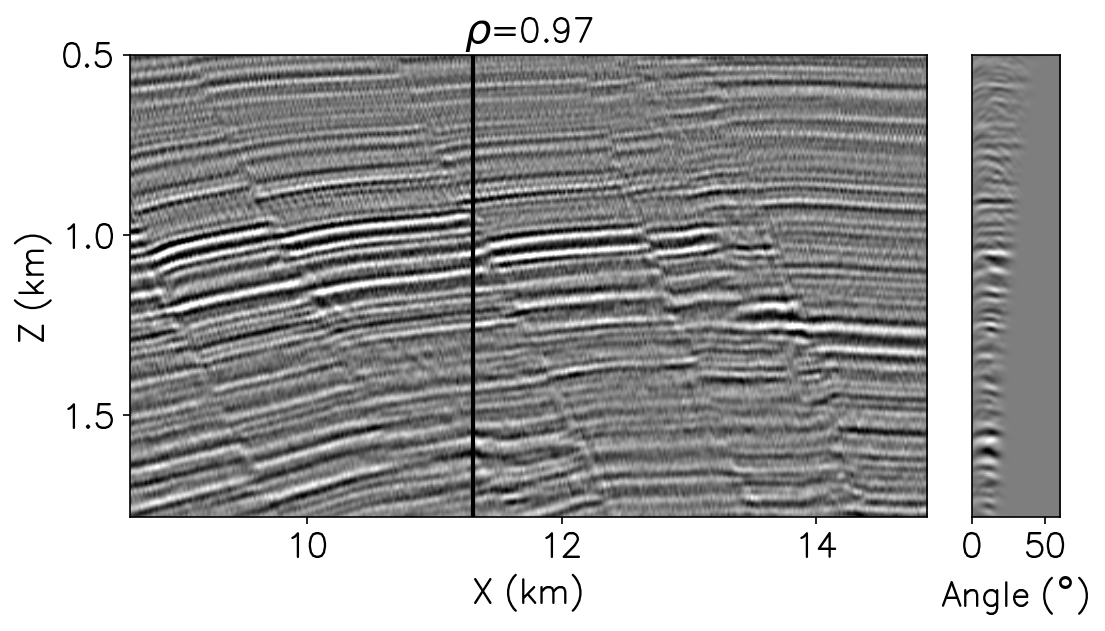} 
  \label{fig:best}}
  \subfigure[]{\includegraphics[width=0.5\textwidth]{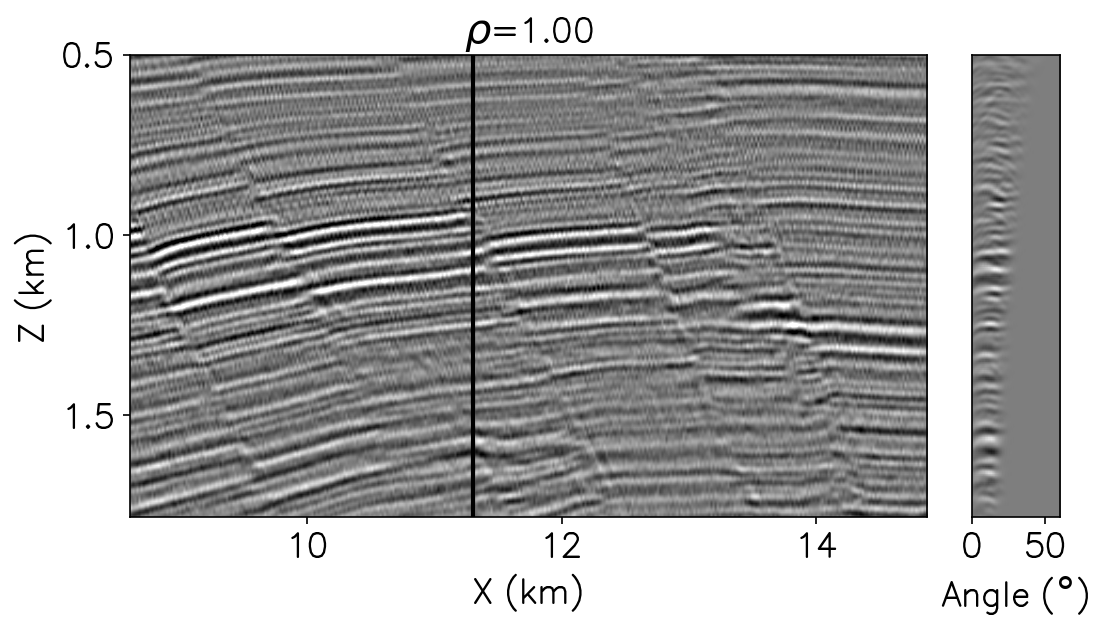}}
  \subfigure[]{\includegraphics[width=0.5\textwidth]{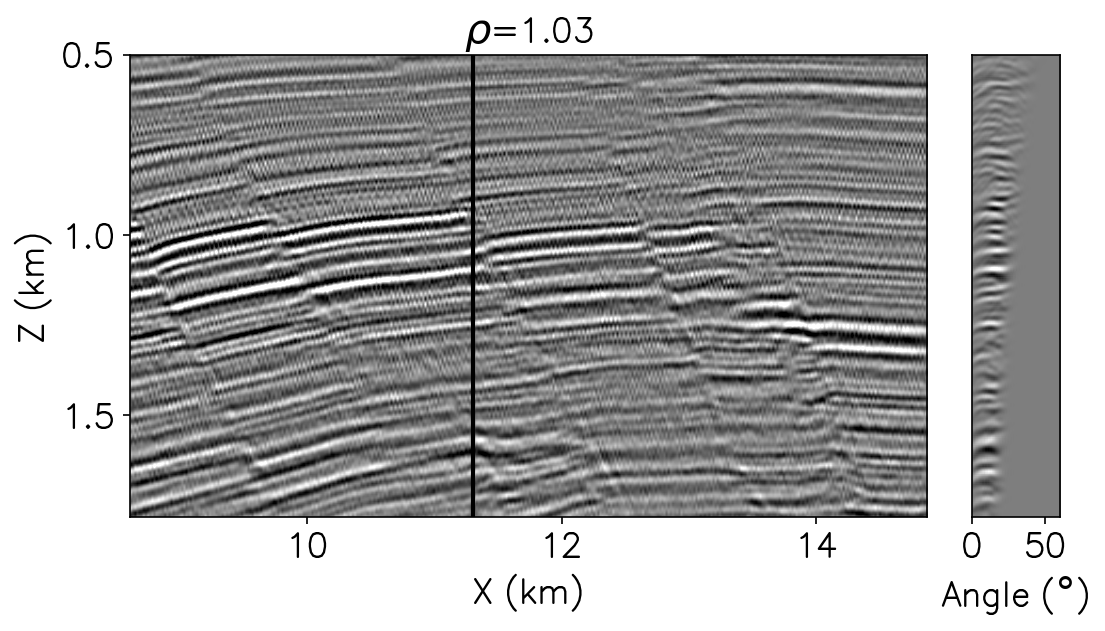}}
  \hspace*{4.0cm}\subfigure[]{\includegraphics[width=0.5\textwidth]{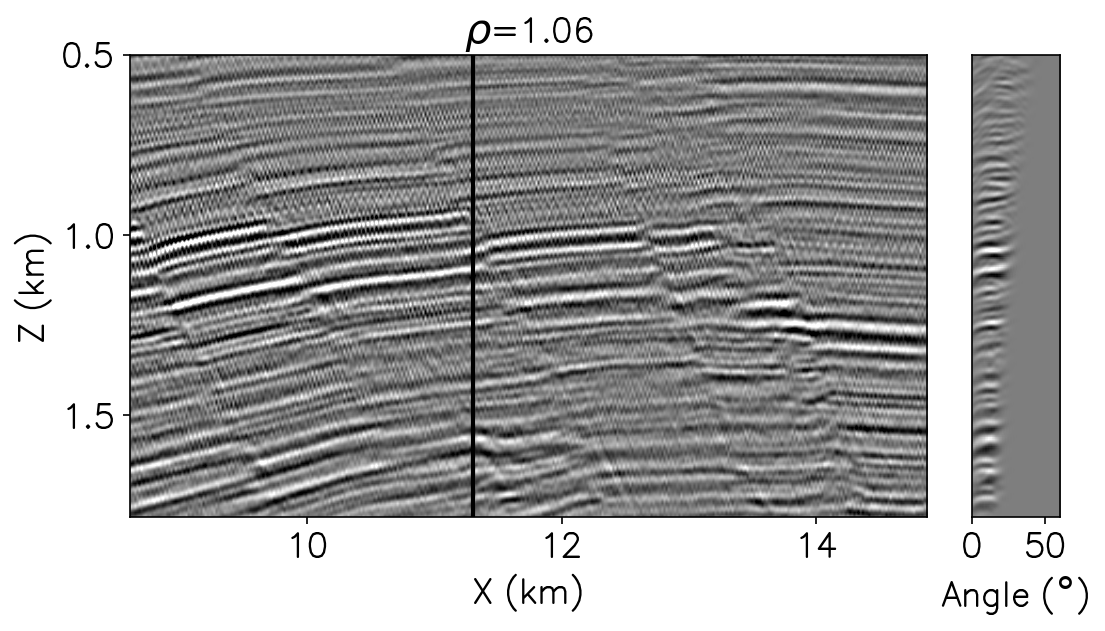}}
  \caption{A selection of residual migration images taken from the application of prestack 
  Stolt residual depth migration to the unfocused image. While it is clear that a single
  $\rho$ value does not fully correct for the velocity error, we observe that the best focusing
  of the faults and largest stack power occur for $\rho=0.97$ (panel (b)).}
  \label{fig:resmig}
\end{figure}

Figure \ref{fig:resmig} shows several images that resulted from applying prestack Stolt 
residual depth migration to the unfocused image shown in Figure \ref{fig:zoomimgmute}. 
We computed residual migration images from $\rho$ values
of 0.90 to 1.1 with an interval of 0.00125 resulting in a total of 161 residually migrated images.
We also applied a depth correction to each of the images in order to correct for 
depth shifts between the different depth migrated images. Additionally, as we did for the
original unfocused image shown in Figure \ref{fig:zoomimgmute}, we converted each 
residually migrated subsurface-offset-domain common-image gather 
to an aperture-angle-domain common-image gather. In Figure \ref{fig:resmig}, 
it is apparent that while none of the images have been fully corrected 
for the velocity error, the image corresponding to $\rho=0.97$ (Figure \ref{fig:best}) 
provides the best focusing of the fault plane reflections as well as the largest stack power. 
Due to the fact that the trial velocity must be a scaled version of the migration velocity, 
only a global refocusing can be performed on the
entire image and the resulting images will at best contain locally-focused regions. 
In order to fully correct for local velocity errors, we must establish focusing criteria 
that can be computed over the entire image and provide an estimate of a spatially-variant 
residual migration parameter ($\rho(z,x)$). 

\subsection*{Measuring image focusing with semblance}
\begin{figure}
  \centering
  \subfigure[]{\includegraphics[width=\textwidth]{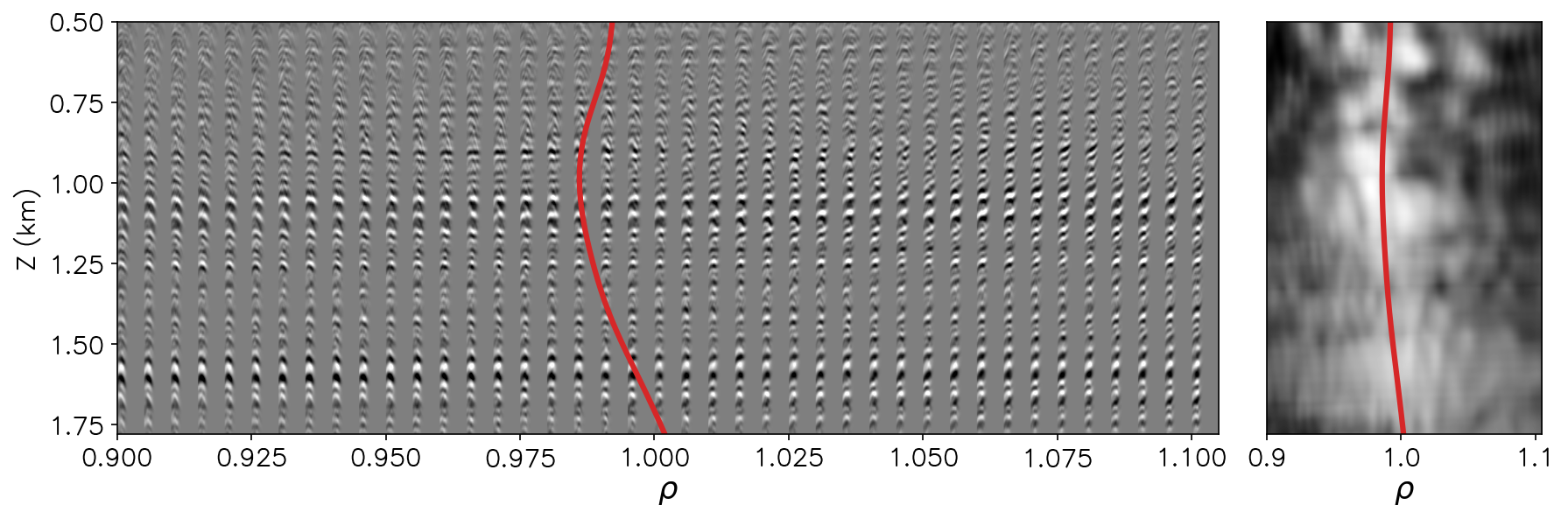} 
  \label{fig:sembmute}}
  \subfigure[]{\includegraphics[width=0.4\textwidth]{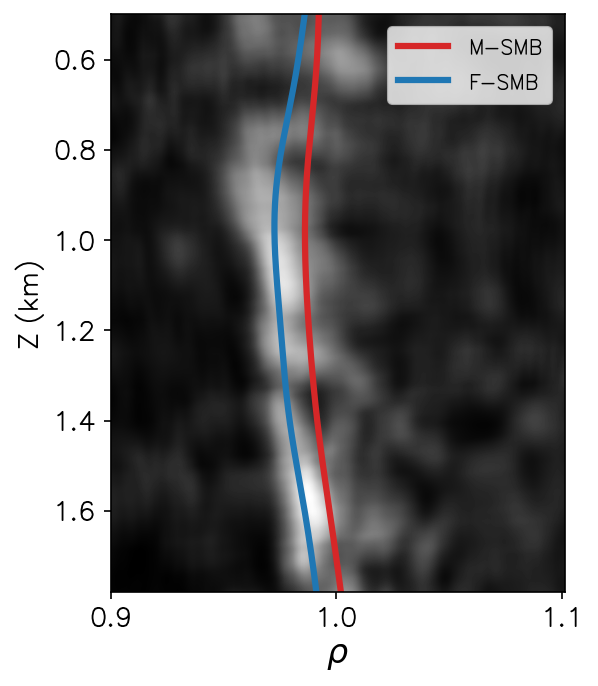} 
  \label{fig:sembfull}}
  \caption{(a) Computed $\rho$ semblance from an angle gather extracted at 11.3 km. 
  The left panel shows
  the angle gather residually migrated for $\rho$ values between 0.9 and 1.1 and
  the right panel shows the computed semblance. 
  The red curves show the pick of the maxima of the semblance panel. 
  (b) The computed $\rho$ semblance from the angle gather extracted at 11.3 km,
  but from the full-aperture image. The blue curve shows the pick of this semblance
  panel (labeled F-SMB) and the red curve is the pick from the limited-aperture
  semblance panel (labeled M-SMB and shown in panel (a)).} 
  \label{fig:semb}
\end{figure}

A simple yet often effective focusing measure is to measure the flatness of the angle gathers. 
This can be done with the following coherence measure \cite[]{neidell1971semblance}:
\begin{equation}
  s_i  = \frac{\sum_{j=i-M}^{i+M} 
  \left(\sum_{k=1}^{N} g_{j,k}\right)^2}{N\sum_{j=i-M}^{i+M}\sum_{k=1}^{N}g_{j,k}^2},
  \label{eqn:semb}
\end{equation}
where $g$ is a residually migrated angle gather for a single $\rho$, 
$i$ and $j$ are depth indices, $k$ is the current trace 
index within the gather, $N$ is the number of traces
within the gather and $M$ is the length of the smoothing filter in depth. 
Performing this calculation for all depths and a range $\rho$ values will result in a 
semblance panel that will indicate, for all depths, the $\rho$ values that provide the flattest 
gathers. The maxima of this 
semblance panel can then be picked resulting in an estimate of $\rho(z)$ for the position 
at which the angle gather was extracted. Figure \ref{fig:sembmute}
shows the result of computing a semblance panel for $\rho=0.9 - 1.1$ for an angle gather 
extracted at 11.3 km within the
unfocused image. We picked the maxima using the automatic picking algorithm 
described in \cite{fomel2009velocity}
and the picks are displayed in the red curves superimposed on the 
angle gather and semblance panels shown in Figure \ref{fig:sembmute}.

While semblance provides an efficient measure for image focusing and in many cases yields 
a robust estimate for $\rho(z,x)$, the resolution of semblance is 
heavily reliant on a sufficient range 
of aperture angles. With a small range of aperture angles, 
the power of the stack will be less sensitive 
to changes in $\rho$, which will result in a broader range of maxima 
in the semblance panel and consequently
less accurate semblance picks. 
To illustrate this, we performed a reference focusing analysis on the 
full aperture image. The semblance computed from the angle gather extracted from the same spatial
location within the full aperture image is shown in Figure \ref{fig:sembfull}.
While we see a tight and clear trend in the semblance panel in Figure \ref{fig:sembfull}, 
we see a very broad smeared trend in the semblance panel in Figure \ref{fig:sembmute}. 
The picks of each of the semblance
panels is also shown in Figure \ref{fig:sembfull}. 
We observe a significant difference between the 
picks from the full and limited-aperture data indicating an inaccurate pick resulted from the 
focusing analysis on the limited-aperture data.

\subsection*{Measuring image focusing with a CNN}

We can overcome this explicit reliance on the aperture angle range 
by augmenting the semblance-based 
focusing analysis with the focusing information contained within the physical axes of the image.
We propose to do this with a 
CNN. In recent years, CNNs have shown remarkable abilities at automating a wide range of
computer vision tasks \cite[]{krizhevsky2012imagenet,taigman2014deepface,liu2016ssd}. 
They have been especially useful for tasks that are difficult
to define formally (easy for humans to perform but difficult to automate with an algorithm). 
Even more recently, CNNs have shown to be of use within the seismic interpretation community. 
Their powerful ability to 
learn to extract features within seismic images has allowed them to detect subsurface structures
in a wide variety of complex geological scenarios.
We extend this idea of geological feature detection to aid in the task of image-focusing analysis.
We do this by training a CNN to classify geological features as focused or unfocused. 
We provide prestack image patches to the CNN and train it to provide a score valued from 
zero to one of the focusing of the patch.
In a similar way to semblance, this focusing score serves as a measure for the
focusing of the patch. In the following sections, we describe the design and 
training of our image-focusing CNN.

\subsubsection*{Neural network structure}

\begin{figure}
  \includegraphics[width=\textwidth]{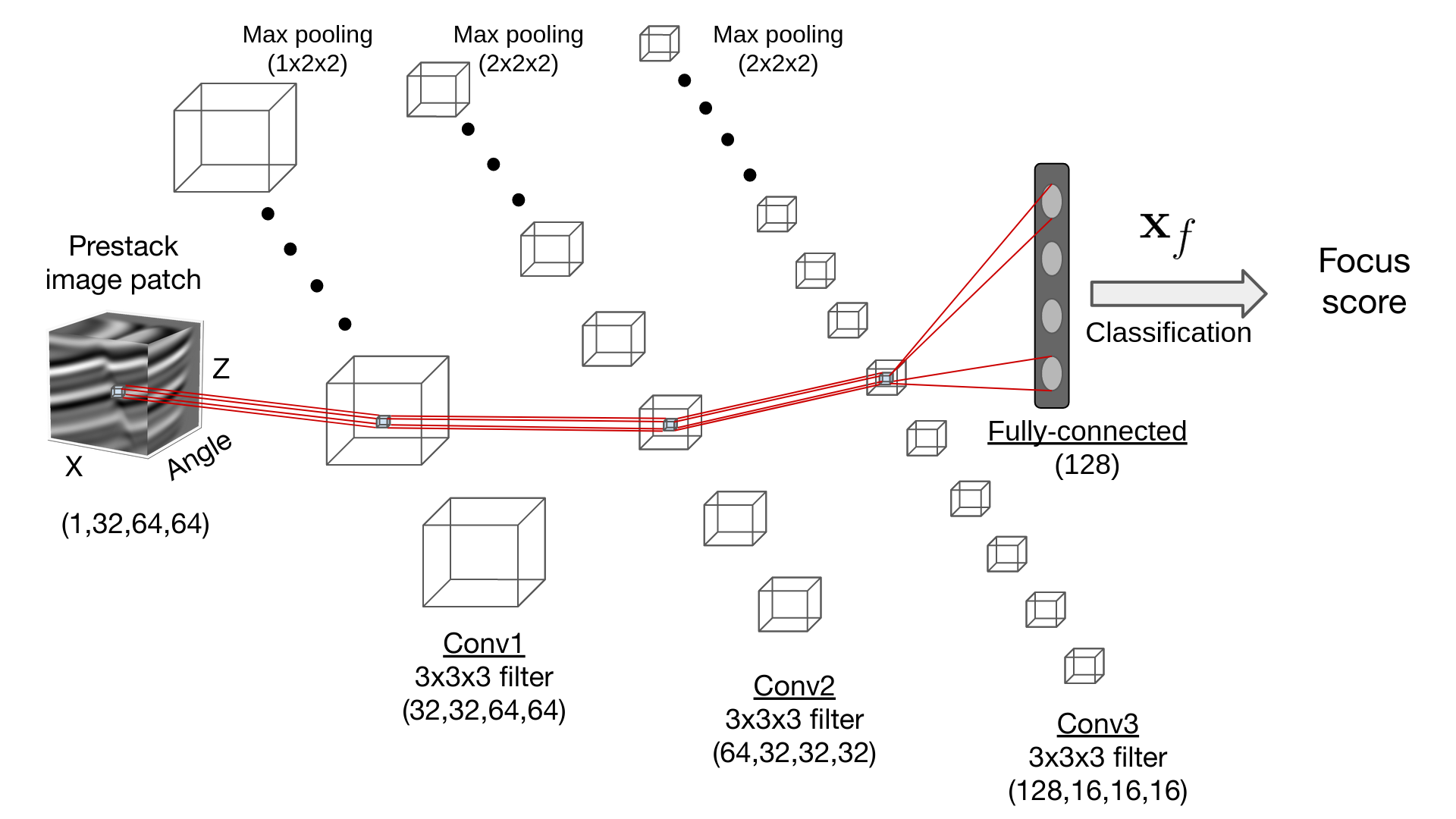}
  \caption{A schematic of the architecture of the feature extraction component 
  our image-focusing CNN. The shapes written beneath each of the operations 
  denote the shapes of the outputs of the operation.}
  \label{fig:cnnarch}
\end{figure}

We designed our CNN to have a feed-forward architecture that is composed of two primary
components: a feature extraction component and a classification component. The feature
extraction component takes as input a prestack image patch (which in 2D has dimensions of
($n\gamma$, $nz$, $nx$), $\gamma$ representing the aperture angle) and provides as output
a feature vector consisting of 128 elements.
Mathematically, we can describe this component as the following non-linear transformation:
\begin{equation}
  \mathbf{x}_f = \mathbf{f}(\mathbf{r}; \mathbf{w}, \mathbf{b}),
\end{equation}
where $\mathbf{x}_f$ is the output feature vector of length 128, $\mathbf{r}$ 
is an input prestack image patch of size $32 \times 64 \times 64$, 
and $\mathbf{w}$ and $\mathbf{b}$ are vectors that contain
the weights (i.e, CNN filter coefficients) and biases to be estimated and that parameterize
the CNN. 
Figure \ref{fig:cnnarch} shows a schematic of the architecture of the CNN.
The network consists of three 3D convolutional, max-pooling and rectified linear unit
(ReLU) blocks. All convolutional operations were performed using filters of size 
$3 \times 3 \times 3$ and apart from the first layer, all max-pooling operations were performed
with a kernel size of $2 \times 2 \times 2$ which act to halve each dimension of the input
features. (For the first block, we chose to not perform pooling along the angle axis).
These three blocks serve to extract the features from the image that are most relevant
for determining the focusing of the image patch.
The output of these blocks is then 
reshaped and transformed into the output feature vector $\mathbf{x}_f$ 
via a fully-connected layer.
The final classification step maps $\mathbf{x}_f$ to a scalar via an 
inner-product and then a sigmoid activation function is used to provide an output score:
\begin{equation}
  s(\mathbf{r}; \mathbf{w},\mathbf{b}, \bm{\theta},\theta_0) = 
  \frac{1}{1 + e^{-\bm{\theta}^T\mathbf{x}_f(\mathbf{r}; \mathbf{w}, \mathbf{b}) + \theta_0}},
\end{equation}
where $\bm{\theta}$ is a 128-element vector of parameters and $\theta_0$ is a bias parameter
to be estimated during training. $s$ is then the output image-focus score (bounded between
0 and 1) that we use as a measure for the focusing of the image patch.

We chose the design of the neural network wanting to take advantage of all of the information
provided within the prestack image patch.
We found that we could best do this by extracting three-dimensional
features with 3D convolutions and 3D max-pooling operations. While we could add 
layers and complexity to our CNN,
we desired to keep the total number of parameters relatively small.
The feature extraction component consists of a total of 8,666,304 parameters 
(number of parameters in $\mathbf{w}$ and $\mathbf{b}$). 
Including the 129 parameters (128 weights from the final classification layer 
($\bm{\theta}$) and the bias term), the CNN has a total of 8,666,433 trainable parameters.

\subsubsection*{Training data}

To create training images to train the CNN, we first formed patches from the limited-aperture
unfocused image. We chose a patch size of 64 depth samples $\times$ 64 lateral samples
and for each patch we selected all 32 angles. We chose to have 50\% overlap in both $z$ and 
$x$ directions when forming patches which resulted in 225 spatial patches. 
After forming patches on all 161 residually-migrated limited-aperture images, 
we obtained a total of $225 \times 161 =$  36,225 patches.

\begin{figure}
  \subfigure[]{\includegraphics[width=0.5\textwidth]{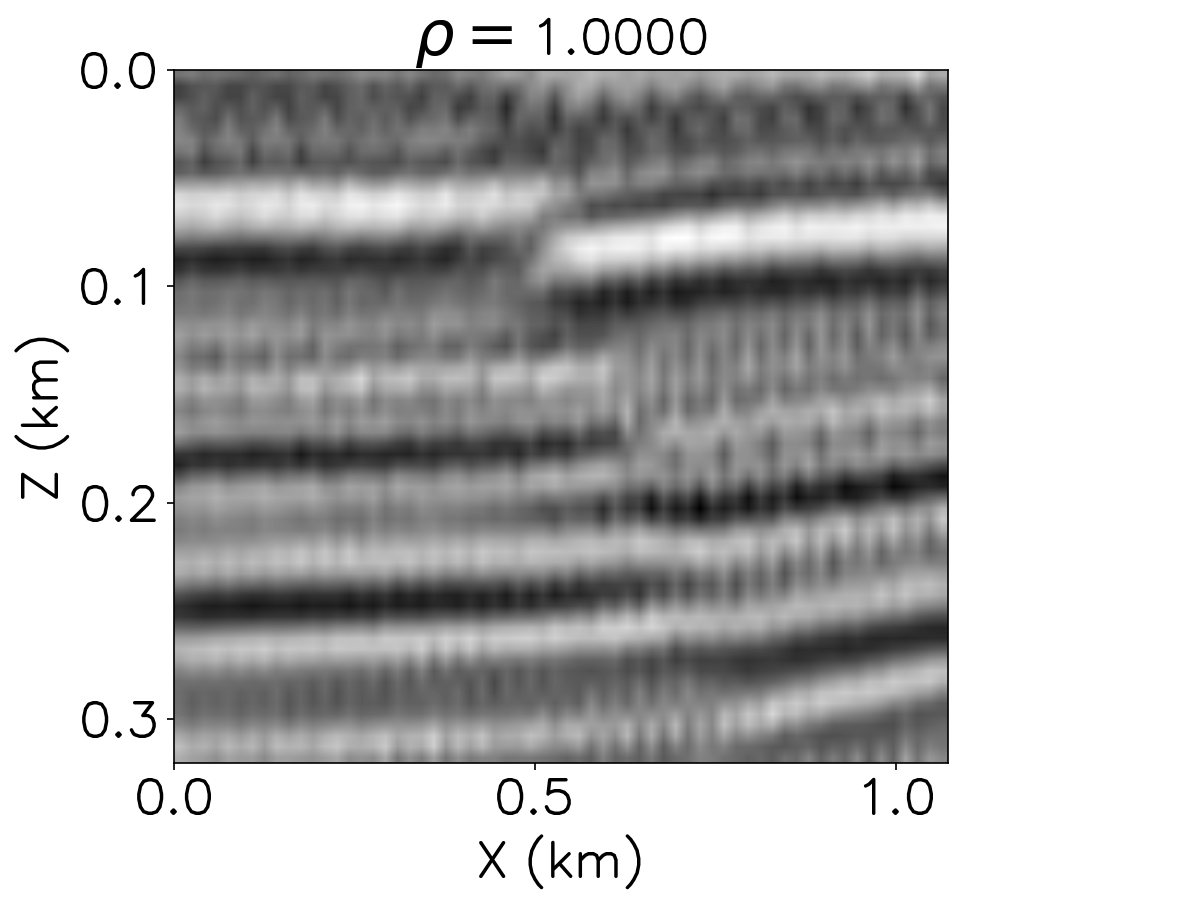}}
  \subfigure[]{\includegraphics[width=0.5\textwidth]{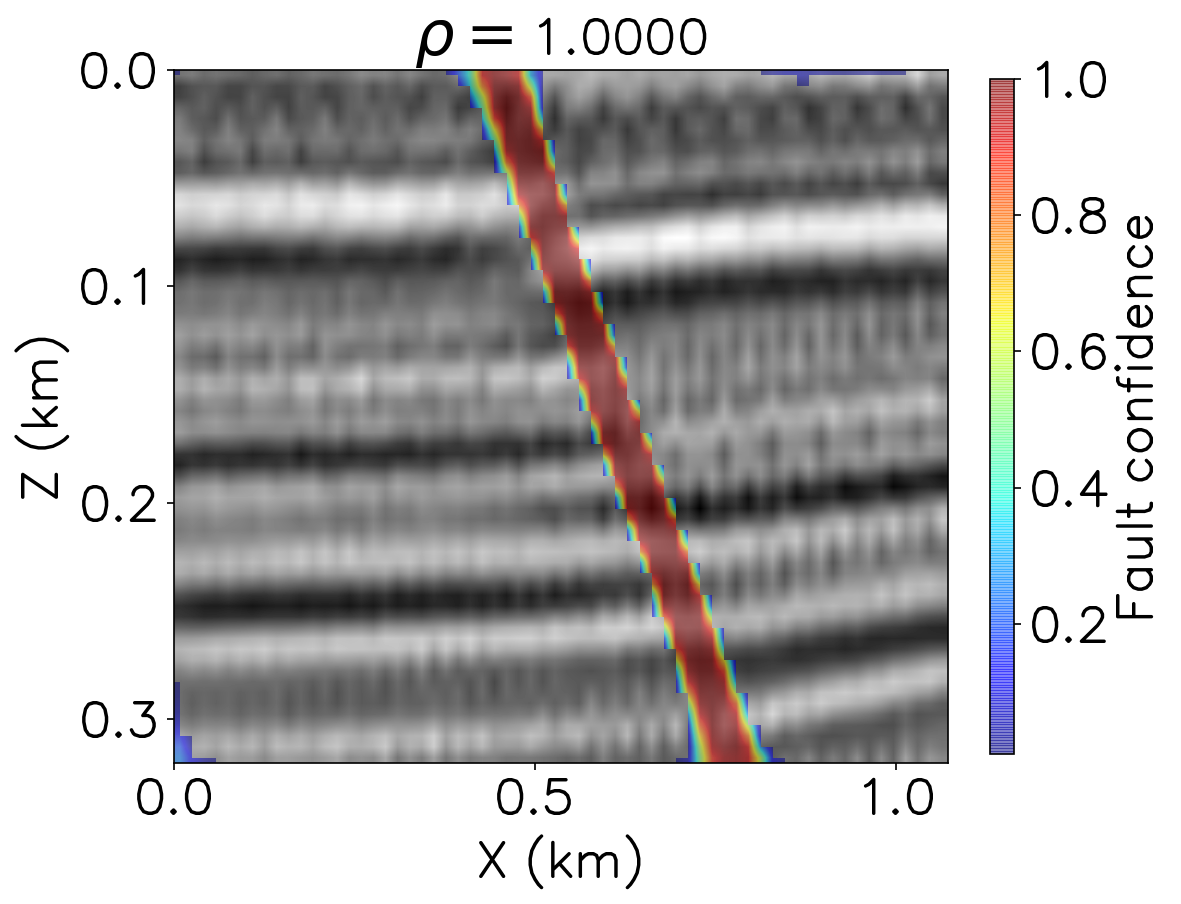} 
  \label{fig:ro1fltseg}}
  \subfigure[]{\includegraphics[width=0.5\textwidth]{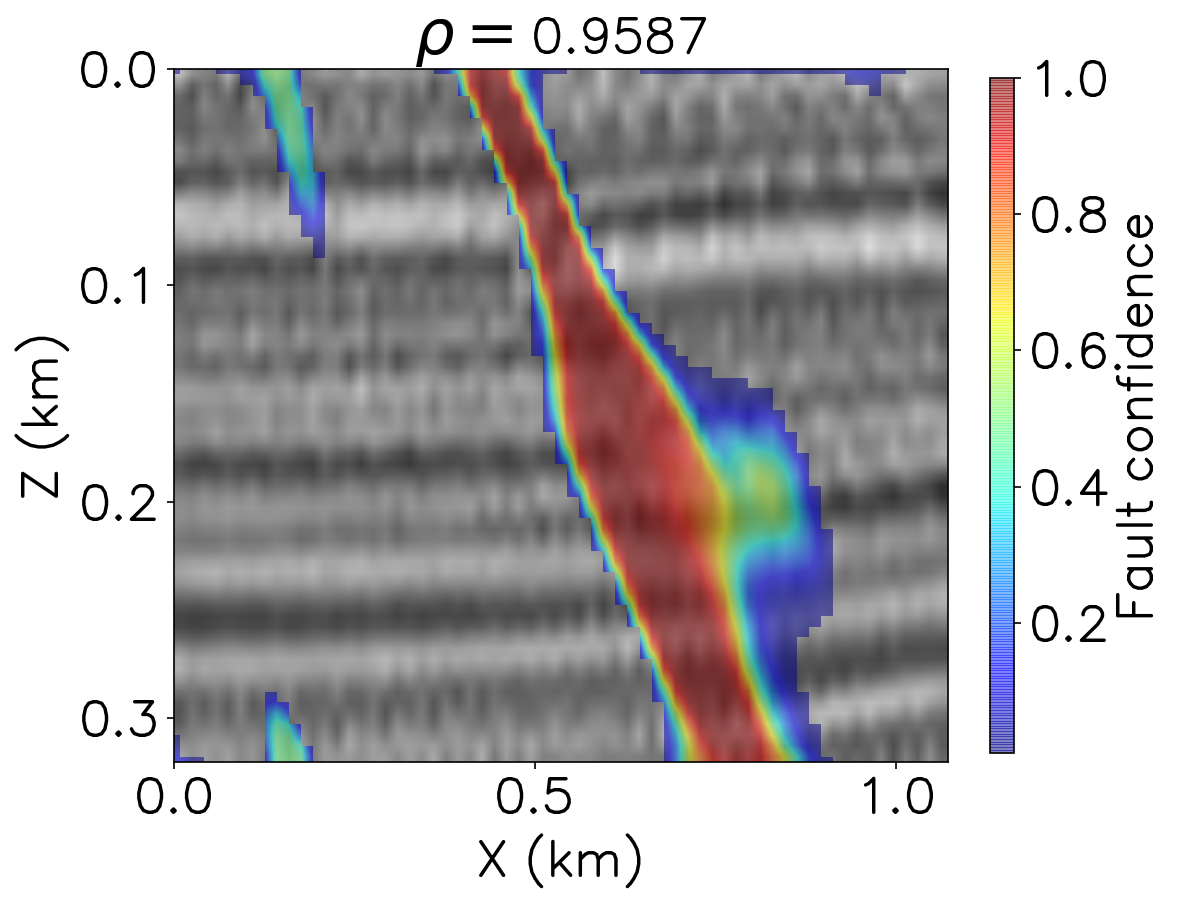}}
  \subfigure[]{\includegraphics[width=0.5\textwidth]{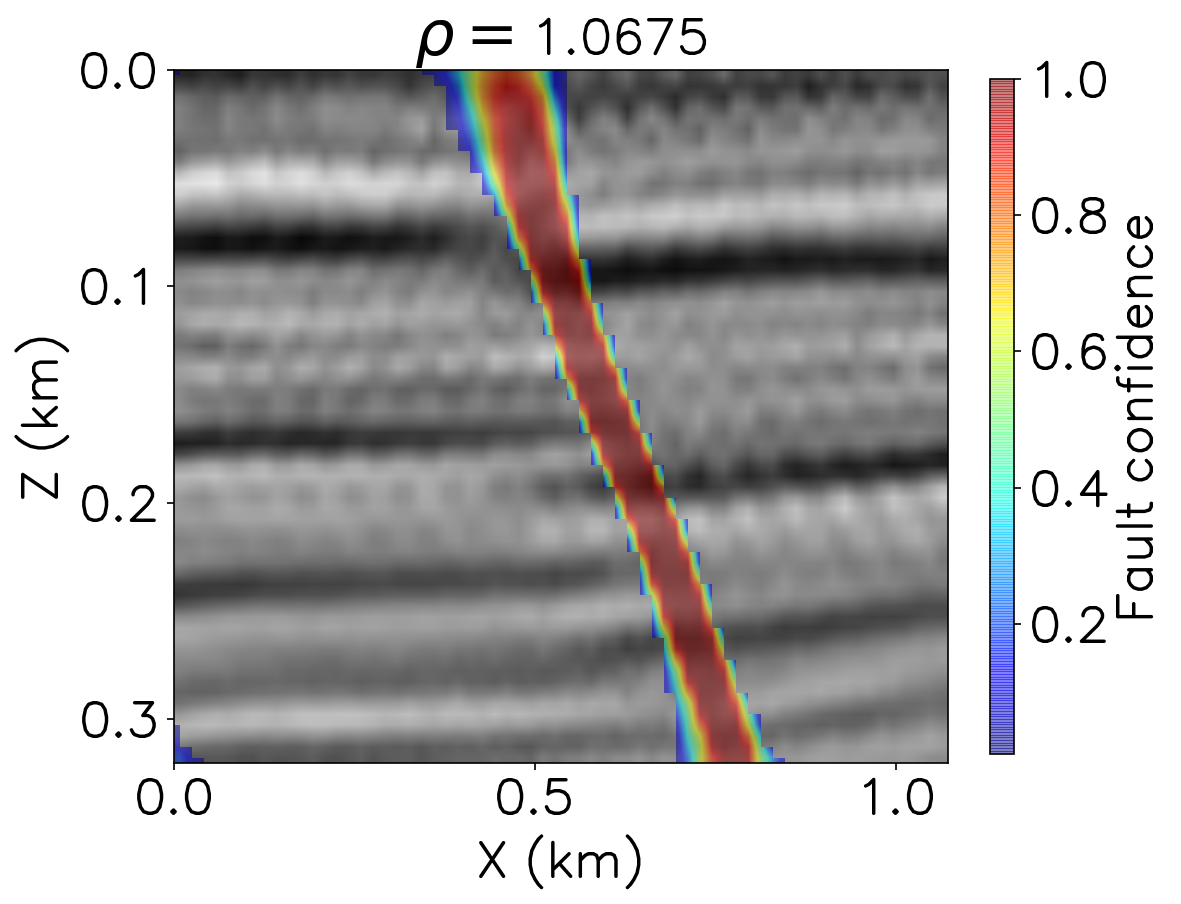}}
  \caption{Fault segmentation on a training patch for different $\rho$ values. Panel (a)
  shows the $\rho=1$ image and panels (b), (c) and (d) show the computed fault confidences
  estimated from the $\rho=1.0$, $\rho=0.9587$ and $\rho=1.0675$ images respectively.
  Qualitatively, comparing the 
  fault segmentations for the different residually migrated images, we observe that the $\rho=1$
  image has the most precise fault segmentation.}
  \label{fig:faultres}
\end{figure}

To label the patches, we grouped all residually-migrated patches that belonged to the same spatial
location and labeled the ``best focused" patch of this group as a focused patch 
(receiving a label of one)
and then all other patches within the patch group we classified as unfocused 
(receiving a label of zero).
For patches without faults and near the surface that were largely 
unaffected by the aperture angle mute, 
we could use the power of the stack as well as visual reflector coherence 
within the patch as a metric for selecting the best focused patch. 
For all other patches, we first performed automatic fault segmentation 
on that patch for the $\rho=1$ image.
For fault segmentation, we used a 2D CNN-based fault segmentation approach trained on 2D
synthetic training images. 
The architecture of our fault segmentation CNN was designed after the U-net
shown in \cite{wu2019faultseg3d}. If an image patch with a segmented
fault contained enough fault pixels indicating that it contained at least one fault, 
we would visually label the patch within the patch group 
that had the most coherent and accurate fault segmentation as focused and all other
patches as unfocused. Figure \ref{fig:faultres} shows the fault
segmentation for a patch for different $\rho$ values.  
Qualitatively, the fault segmentation for the
$\rho=1$ image shown in Figure \ref{fig:ro1fltseg} 
appears to have the sharpest and most precise fault
segmentation and therefore, this image patch received the focused label for this patch group.

\begin{figure}
  \subfigure[]{\includegraphics[width=\textwidth]{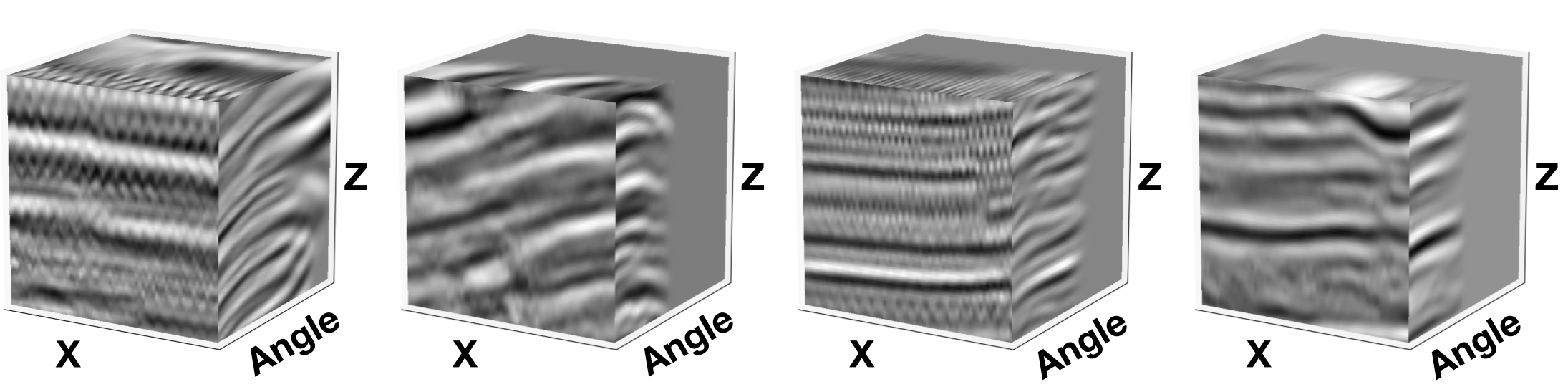} 
  \label{fig:fldpatches}}
  \subfigure[]{\includegraphics[width=\textwidth]{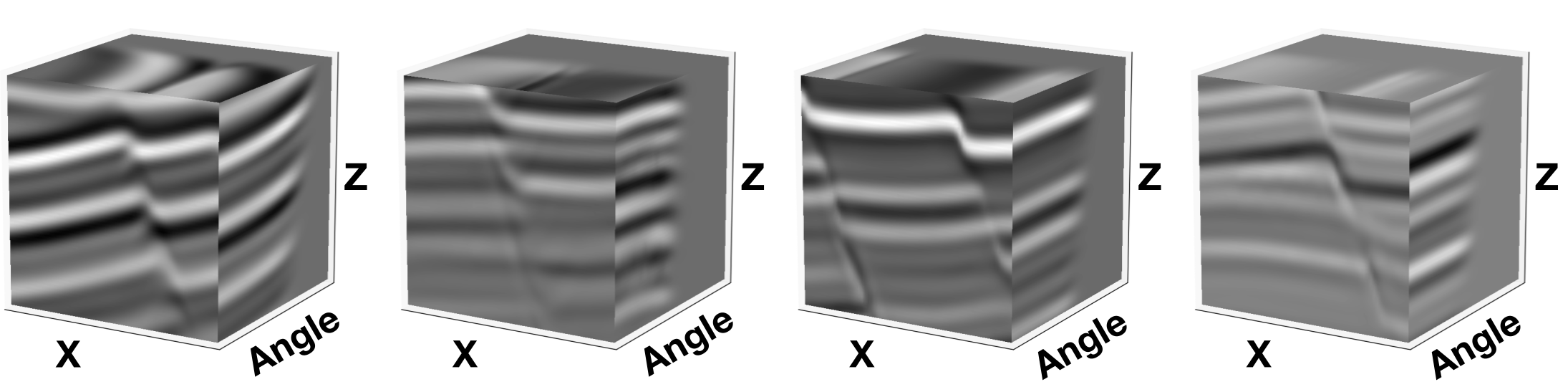} 
  \label{fig:synpatches}}
  \caption{A selection of prestack training image patches used to train our image-focusing CNN.
  (a) Labeled patches taken from outside of the region of interest within 
  the limited-aperture image and  (b) labeled patches created from our
  synthetic training image procedure. For both panels (a) and (b), 
  the two leftmost images are unfocused training patches and the rightmost images 
  are focused training patches.}
  \label{fig:trnpatches}
\end{figure}

While this approach for creating a labeled training set is advantageous 
in that we do not have to be as concerned with model generalization and a data-domain gap, 
the labeling procedure can be time-consuming and also the resulting labeled images
are dependent on the individual performing the labeling. Additionally, this approach 
will generate far fewer focused training images than unfocused training images as only one
out of 161 is labeled focused for each patch group. While this can still result in a 
relatively large number of training patches, it will lead to a largely unbalanced dataset that 
will likely bias the model to predicting unfocused patches. For this reason,
we select only one focused and one unfocused patch for each patch group (the unfocused
patch is selected at random from the unfocused patches in the patch group).
This will lead to a perfectly balanced dataset but has the
downside of providing significantly fewer patches. Due to this reason, as well as our 
desire to not spend too much time on manually labeling patches, 
we labeled only 56 field data patches. Figure \ref{fig:fldpatches} shows examples
of our unfocused and focused field training patches. 
Note that for the purposes of this paper and to 
investigate generalization, all labeled field patches used for training 
were selected from outside of the region of interest.

To overcome this challenge of limited labeled field training patches, we supplemented the field 
training set with a synthetic pre-training set. 
The addition of a synthetic pre-training set also has
the advantage in that it can help reduce biases and uncertainties 
that arise due to a manual labeling procedure.
To create synthetic training image patches we implemented the following procedure: 
we first create a synthetic reflectivity 
model with undulating reflectors as well as normal faults. 
We then convolve this reflectivity model with a 20 Hz Ricker wavelet to 
create a band-limited synthetic seismic image.
To minimize the domain gap between the synthetic and real images, we attempt to
make the geological structure and frequency content of our synthetic seismic image similar to
our real seismic image. We then simulate a subsurface-offset image by padding 
with 20 positive and 20 negative offsets keeping our 
synthetic image at the zero subsurface offset 
(a focused image). To create an unfocused image, we then perform prestack Stolt residual 
depth migration for a random $\rho$ value that is different from $\rho=1$.  
This will create a constant velocity error on the image resulting in diffractions on faults 
and energy away from zero subsurface offset. Finally, we convert both images from the 
subsurface-offset domain to the aperture-angle domain and 
apply a mute on the gathers to mimic the aperture range on our real seismic image. 
For each synthetic seismic image, we then create patches from each image using the same 
patch grid we used for the real seismic image. Performing this procedure for 1,000 models 
created 8,192 unfocused and focused synthetic pre-training patches. Similar to 
Figure \ref{fig:fldpatches}, Figure \ref{fig:synpatches} shows examples of unfocused and focused 
synthetic pre-training patches.

\subsubsection*{Training the image-focusing CNN}

\begin{figure}
  \centering
  \subfigure[]{\includegraphics[width=0.7\textwidth]{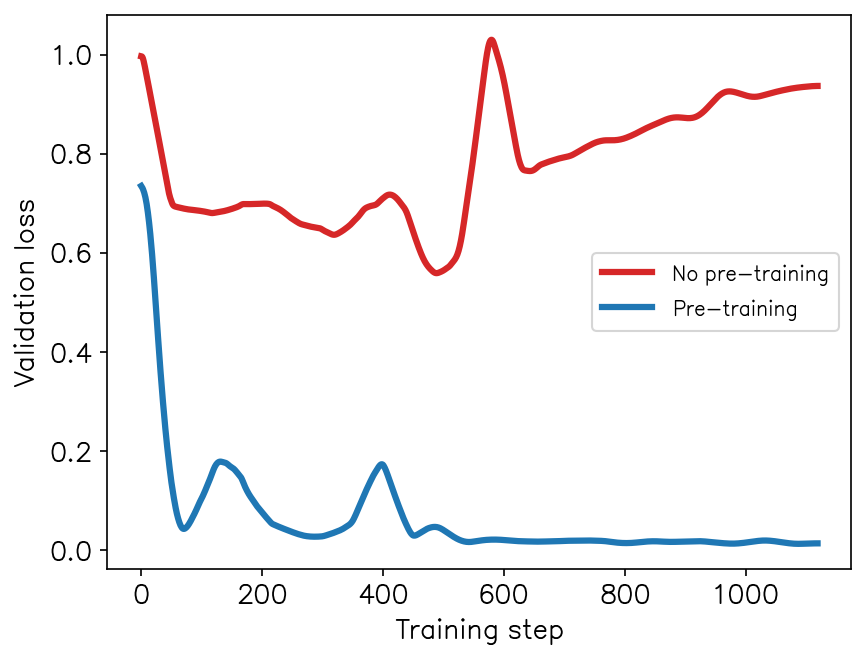}}
  \subfigure[]{\includegraphics[width=0.7\textwidth]{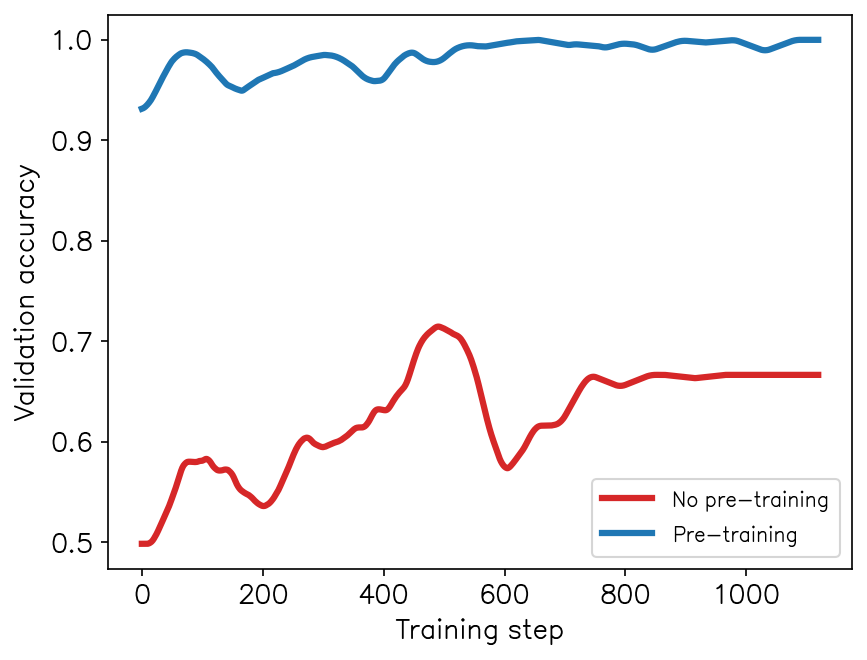}}
  \caption{Learning curves with and without the pre-training stage. Panel (a) shows
  the comparison between the validation loss and panel (b) the comparison of the 
  validation accuracy of the model with and without pre-training. Note to better display
  the trend of the curves, each of the curves was smoothed with a 
  50-point triangular smoothing filter.}
  \label{fig:lcurves}
\end{figure}

To train our CNN, we used the binary cross-entropy loss function:
\begin{equation}
  \mathcal{L}(\mathbf{w}, \mathbf{b}, \bm{\theta},\theta_0) = 
  - \sum_{i=1}^{M} y_i\log(s_i(\mathbf{r}; \mathbf{w}, \mathbf{b}, \bm{\theta}, \theta_0)) + 
  (1-y_i)\log(1-s_i(\mathbf{r}; \mathbf{w}, \mathbf{b}, \bm{\theta}, \theta_0)),
\end{equation}
where $y_i$ is the label for the $i$-th training example and $M$ is the total number 
of examples used for training.
To minimize this loss and estimate the unknown parameters 
$\mathbf{w}, \mathbf{b}, \bm{\theta},$ and $\theta_0$,
we chose the ADAM optimization algorithm with a fixed learning rate of 
$1 \times 10^{-4}$ \cite[]{kingma2014adam}.
The training itself was split into two stages. The first stage was a pre-training stage in which
we trained the CNN on the synthetic pre-training images. For this stage, we specified a
batch size of 20 image patches and trained for 20 epochs and 
used 80\% of the 8,192 patches for training and 20\% for validation. 
For the synthetic validation training set we achieved nearly 100\% accuracy in classification.
In the secondary stage of training we trained on the 56 field training patches
and used an additional 20 for validation. 
We also trained for 20 epochs but used a batch size of one image patch as we found that
smaller batch sizes provided the best validation accuracy. 
We found that with the addition of the pre-training set, 
we were able to achieve very good classification accuracy on the 
validation data as opposed to training with only
the field training patches. 
Figure \ref{fig:lcurves} shows the comparison of the learning curves of the secondary stage
of training with and without the pre-training step. 
We observe that by first pre-training on our synthetic
seismic images, the network was able to avoid overfitting the 
real data training patches and achieve nearly perfect validation accuracy.

\section*{Results}

\begin{figure}
  \centering
  \subfigure[]{\includegraphics[width=0.72\textwidth]{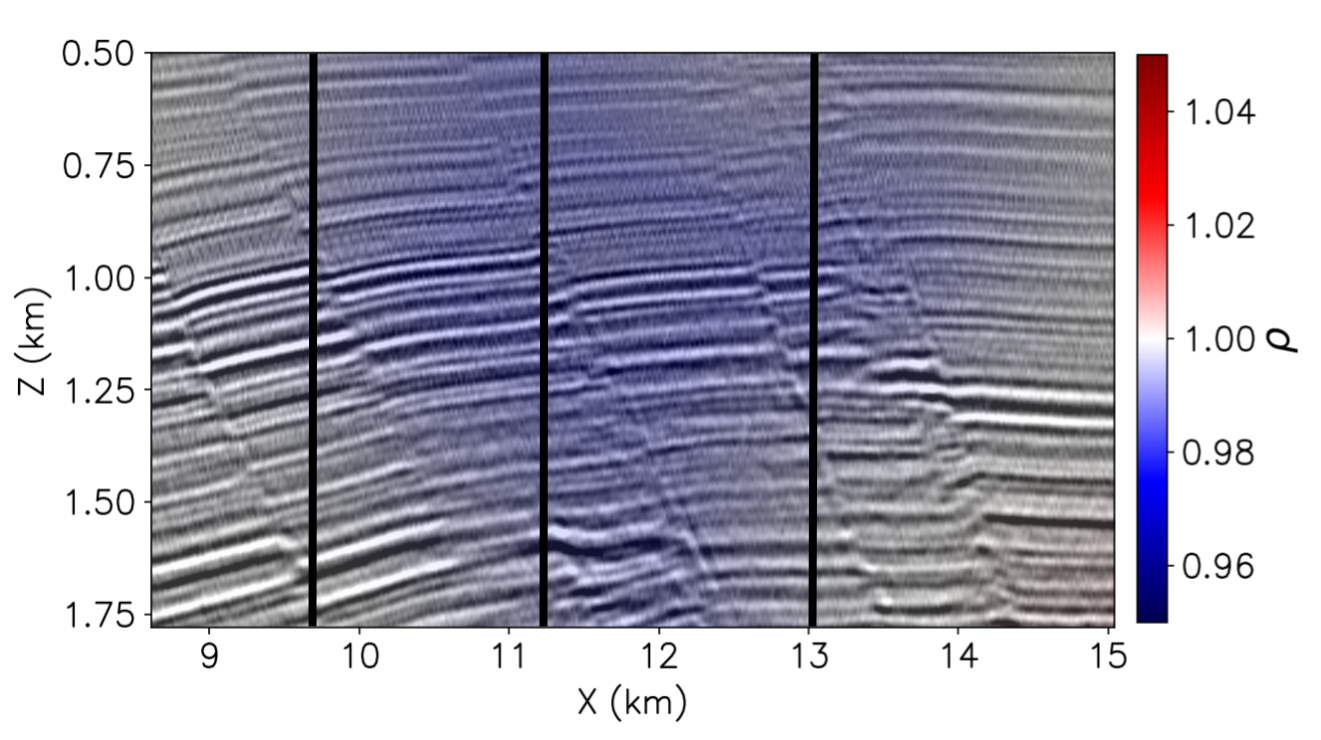} 
  \label{fig:rhocnn}}
  \subfigure[]{\includegraphics[width=0.72\textwidth]{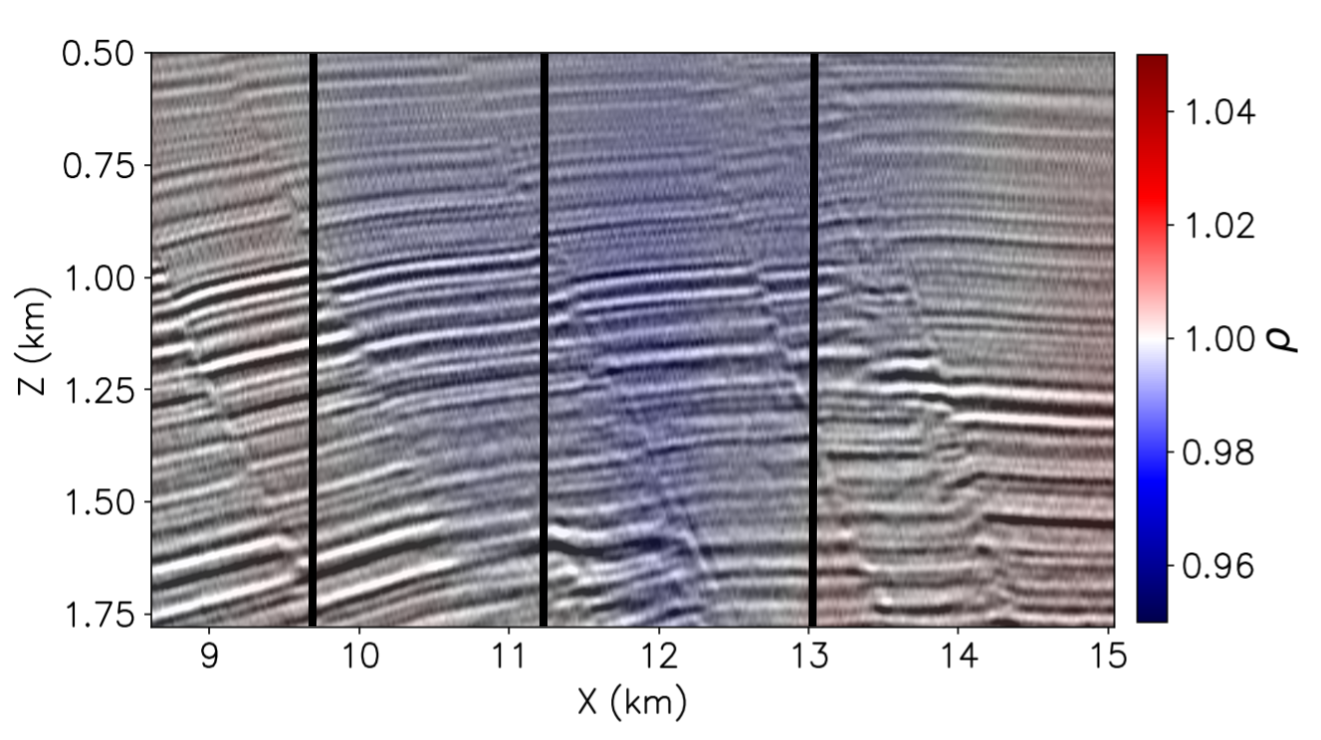} 
  \label{fig:rhosemb}}
  \subfigure[]{\includegraphics[width=0.72\textwidth]{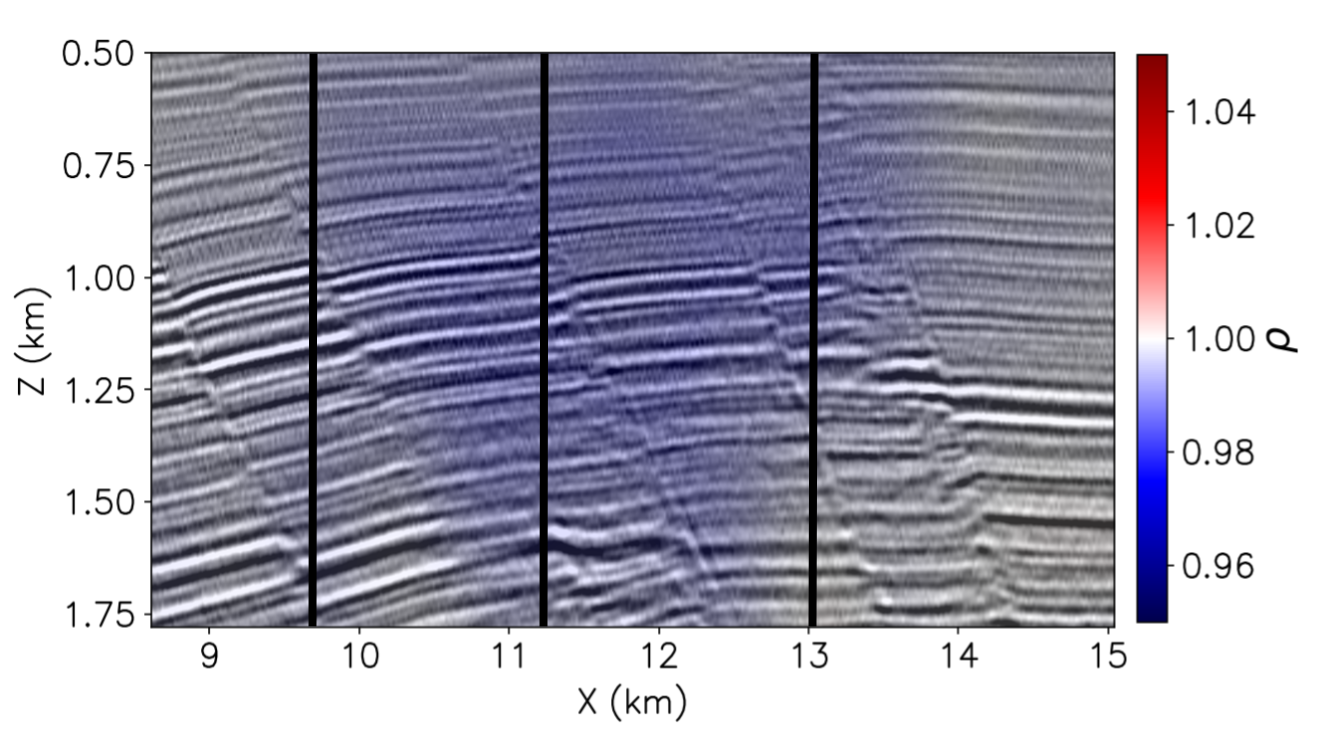} 
  \label{fig:rhoref}}
  \caption{Comparison of estimated $\rho(z, x)$ for (a) our CNN-based approach, (b)
  the semblance-based approach and (c) the semblance-based approach computed
  on the full-aperture dataset included for reference. 
  The black vertical bars indicate the spatial
  locations at which the $\rho(z,x)$ values were extracted and compared in 
  Figure \ref{fig:sembcomp}.}
  \label{fig:rhocomp}
\end{figure}

\begin{figure}
  \centering
  \subfigure[]{\includegraphics[width=0.352\textwidth]{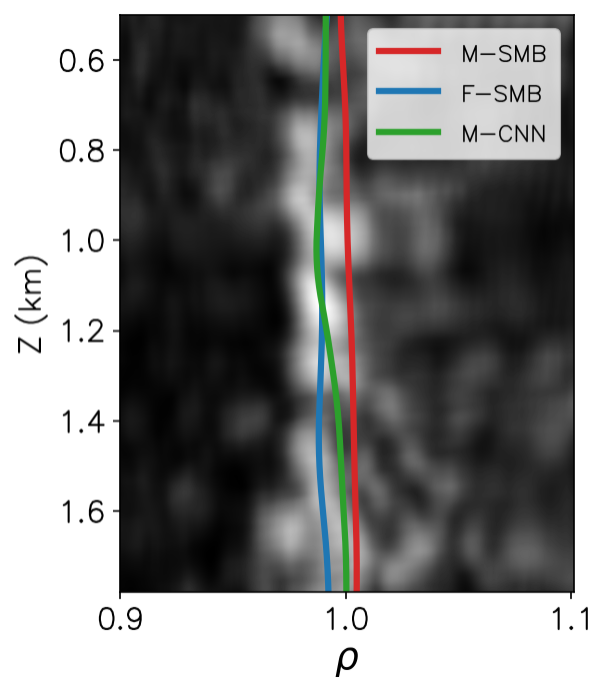}}
  \subfigure[]{\includegraphics[width=0.3\textwidth]{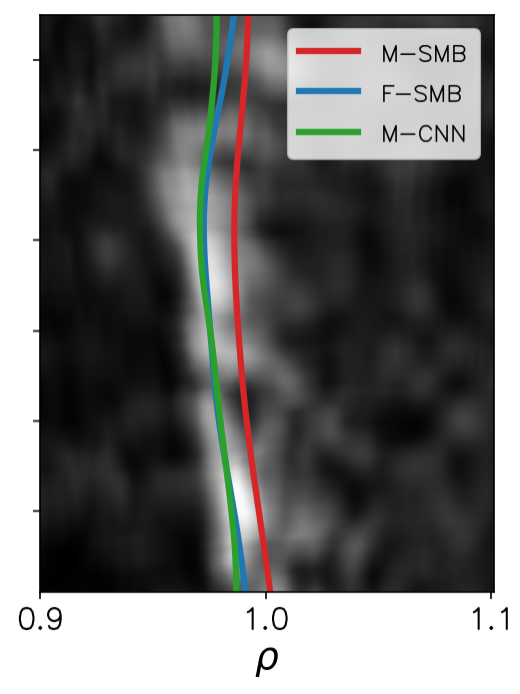}}
  \subfigure[]{\includegraphics[width=0.3\textwidth]{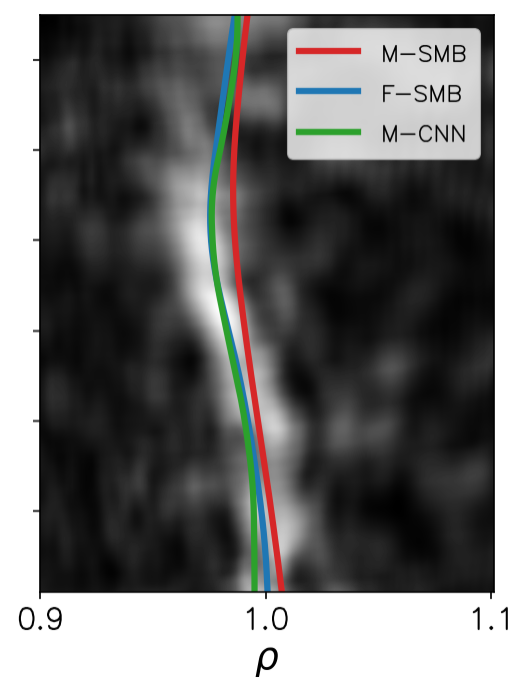}}
  \caption{Comparison of $\rho(z)$ 
  extracted at (a) $x=9.6$ km, (b) $x=11.3$ km and (c) $x=13$ km.  
  Each semblance panel was computed from the reference full-aperture image-focusing analysis.
  The red curve shows the result of performing semblance-based focusing analysis and 
  the green curve the result from the CNN (both performed on the limited-aperture data). The blue
  curve shows the pick of the semblance panels for reference.}
  \label{fig:sembcomp}
\end{figure}

To test the trained image-focusing CNN, we made predictions on image patches
within our region of interest. Forming the region of interest into residually migrated
patch groups, we then computed the image-focusing score for each patch in the patch 
group and chose the $\rho$ value corresponding to the patch that had the maximum focusing
score. We assigned that selected $\rho$ value over the whole patch and performed the same
operation for all patches within the target region. We then smoothed the selected $\rho$ values
with a 0.5 km triangular smoothing filter along the lateral direction and a 0.15 km 
triangular smoothing filter in depth.
Figure \ref{fig:rhocnn} shows the predicted $\rho(z,x)$ 
superimposed on the unfocused image. We observe that
the predicted $\rho(z,x)$ smoothly varies in the unfocused region between 
values of 0.96 and 0.98 in the region
of most severe defocusing which is consistent with what we observed in the 
residual migration images 
(Figure \ref{fig:resmig}). 
Figure \ref{fig:rhosemb} shows the estimated $\rho(z,x)$ 
from the semblance-based approach. While
the two predicted $\rho(z,x)$ generally agree spatially, 
we observe significant differences in magnitude in the predicted
residual migration parameter. 
Additionally, we observe that the estimated $\rho(z,x)$ from the semblance-based
approach oscillates above and below $\rho=1$. 
In order to better compare the estimated $\rho(z,x)$, we performed
a semblance-based focusing analysis for all spatial locations within the 
full-aperture image. The results of this focusing analysis are shown in Figure 
\ref{fig:rhoref}. 
Comparing all three estimated $\rho(z,x)$, we observe both spatially and in terms of magnitude
that the CNN-based approach agrees quite well with the full aperture semblance-based 
focusing analysis. 
Figure \ref{fig:sembcomp} shows a comparison of the estimated $\rho(z,x)$ 
for all three focusing analyses,
taken at $x = $ 9.6, 11.3 and 13 km. For both panels (b) and (c) of 
Figure \ref{fig:sembcomp}, we observe near
perfect agreement between the full-aperture analysis and CNN analysis. 
We also observe good agreement
in panel (a), but we observe that the $\rho(z)$
from the CNN-based approach diverges from the full-aperture semblance approach below
a depth of about 1.2 km.

\subsection*{Quality check of the results}
\begin{figure}
  \centering
  \subfigure[]{\includegraphics[width=\textwidth]{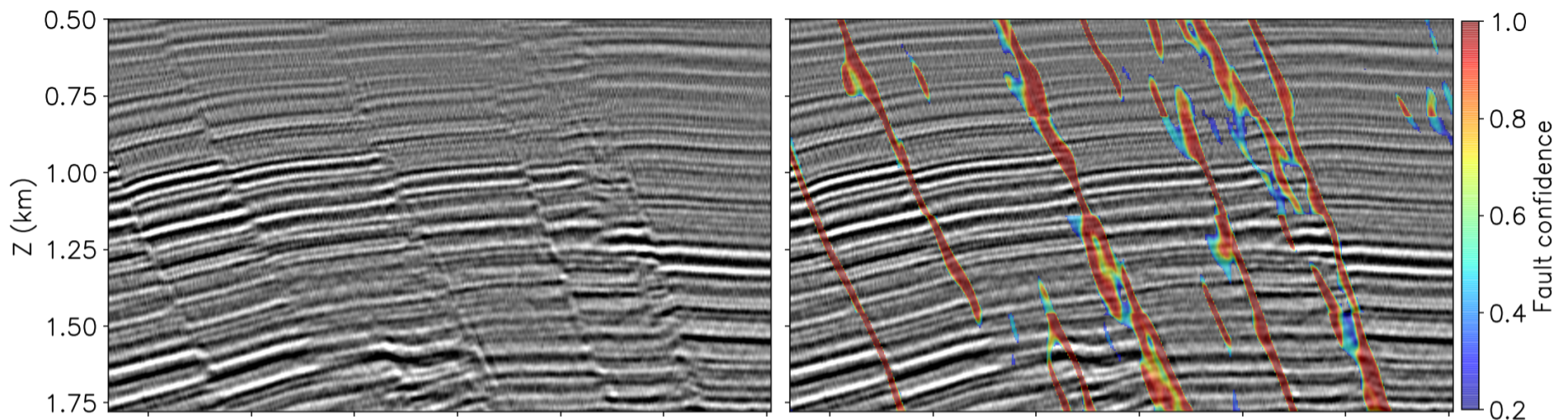} 
  \label{fig:refunf}}
  \subfigure[]{\includegraphics[width=\textwidth]{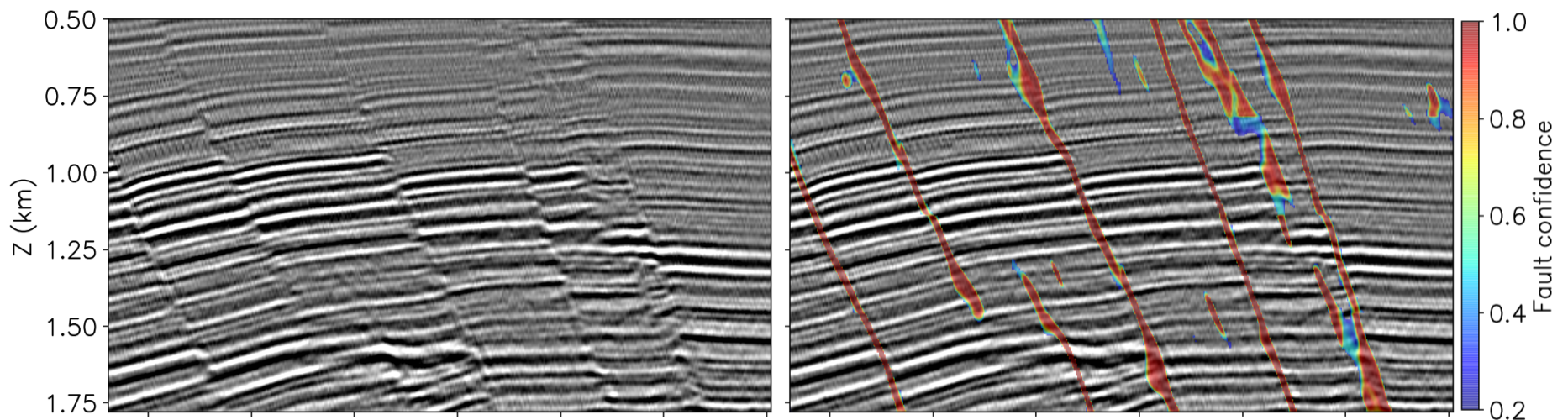} 
  \label{fig:refcnn}}
  \subfigure[]{\includegraphics[width=\textwidth]{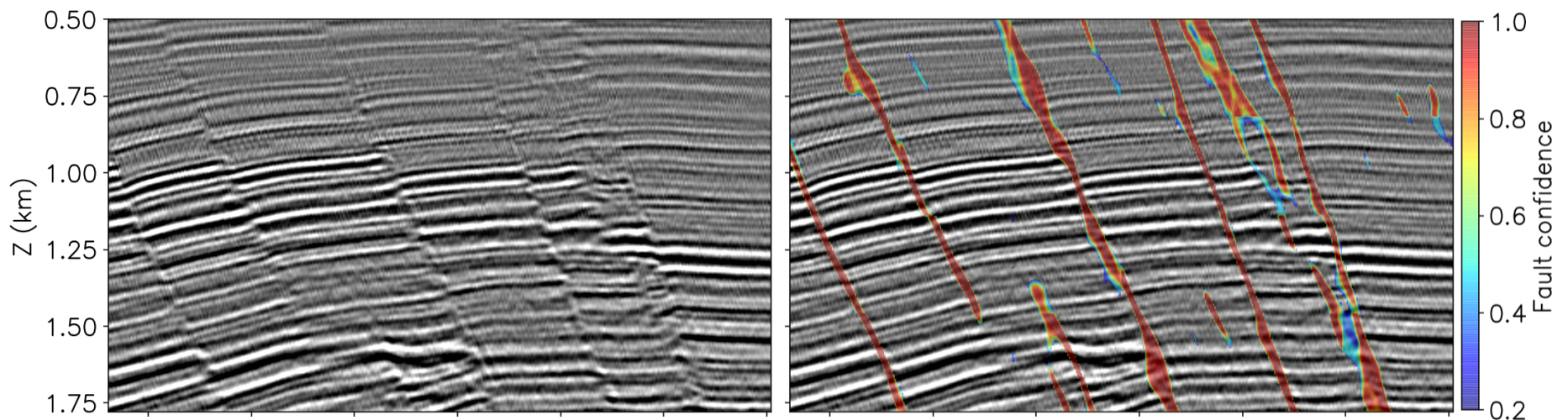} 
  \label{fig:refsmb}}
  \subfigure[]{\includegraphics[width=\textwidth]{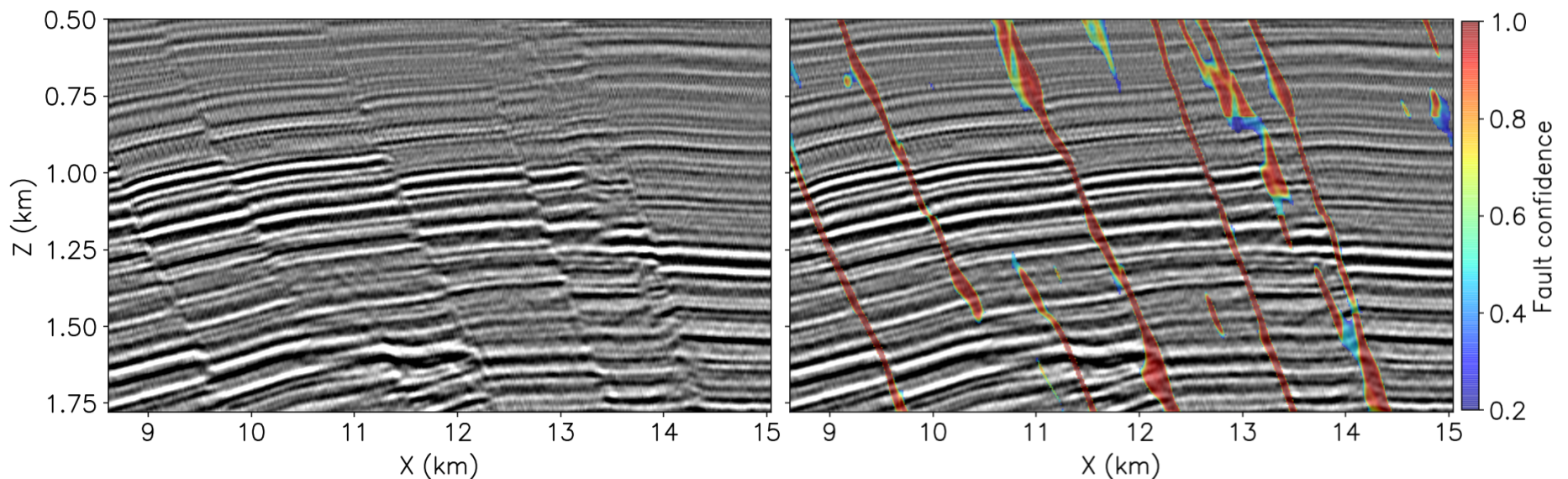} 
  \label{fig:refref}}
  \caption{Comparison of corrected images and fault segmentation. The left column shows the image
  and the right column is the image with the fault segmentation of 
  (a) the original unfocused image, (b) the image corrected with the CNN-based approach, 
  (c) the image corrected with semblance and (d) the reference image
  corrected from full-aperture focusing analysis.}
  \label{fig:refsegcomp}
\end{figure}

In addition to comparing the estimated $\rho$ from the limited-aperture focusing analyses with the
reference full aperture focusing analysis, we performed an additional quality check (QC) to assess
the quality of the estimated $\rho(z,x)$. Our additional QC consisted of first performing a
correction to the unfocused image that extracts the optimally focused 
residual migration images from the collection of stacked residually
migrated images $r(z,x,\rho)$ using the estimated $\rho(z,x)$ \cite[]{mackay1989refining}. 
Mathematically we can express this as
follows:
\begin{equation}
  r_{\text{corr}}(z,x) = r(z, x, \rho(z, x)),
\end{equation}
where $r_\text{corr}$ is the corrected image.
We numerically implement this operation using a high-order spline interpolation.
Panels (b)-(d) of the left column of Figure \ref{fig:refsegcomp} show the results of performing
this correction using the estimated migration scanning parameters
from our CNN-based approach computed from the limited-aperture image, 
the semblance-based approach computed on the limited-aperture image and 
the reference semblance-based focusing analysis computed on the full-aperture image.
Comparing these panels with the original unfocused image in the left column of panel (a)
of Figure \ref{fig:refsegcomp}, 
we observe significant improvement in the focusing
of the faults for each of the corrected images. However, we observe some residual
defocusing in the muted semblance image (Figure \ref{fig:refsmb}). 
This is more readily apparent in the fault segmentation of 
each image shown in the right column of Figure \ref{fig:refsegcomp}. 
Qualitatively, comparing the central fault of 
the fault segmented images on panels (b)-(d), 
we observe at approximately 0.8 km depth that the segmentation
of panels (b) and (d) is better resolved than that of panel (c). 

To provide a quantitative comparison of the
observed improvement in the fault segmentation, 
we first specified the faults segmented on the reference image 
(Figure \ref{fig:refref}) as the ground truth fault locations and applied a threshold to the 
predicted fault confidences in each of the images
where any pixel containing less than 0.3 fault confidence was 
set to zero and all others received a value of one.
We then used a standard intersection over union (IOU) metric for 
assessing the quality of the predicted faults in
the images shown in Figures \ref{fig:refunf} - \ref{fig:refsmb}. 
The IOU metric can be computed as follows:
\begin{equation}
  \text{IOU} = \frac{N_{11}}{N_{11} + N_{01} + N_{10}},
\end{equation}
where $N_{11}$ is the number of fault pixels within the predicted image that matched 
with the fault pixels
in the reference image (true positives), 
$N_{01}$ is the number of predicted non-fault pixels that were actually fault pixels
in the reference image (false negatives) 
and $N_{10}$ is the number of predicted fault pixels that were 
actually non-fault pixels in the target image (false positives).
Table \ref{tab:iou} shows
the computed IOU values from the images shown in Figures \ref{fig:refunf}-\ref{fig:refsmb}. 
We observe a significant improvement in the IOU of the faults segmented within the 
CNN corrected image compared to both the original unfocused and semblance corrected image.

\begin{table}
  \centering
  \begin{tabular}{ |c|c|c|c| }
    \hline
    & \textbf{Unfocused} & \textbf{Semblance} & \textbf{CNN} \\
    \hline
    \textbf{IOU} & 0.58 & 0.72 & 0.83 \\
    \hline
  \end{tabular}
  \caption{Computed IOU metrics for the faults segmented within the original unfocused
  image (Figure \ref{fig:refunf}), the image corrected after semblance-based focusing analysis
  (Figure \ref{fig:refsmb}) and the image corrected after our CNN-based focusing analysis
  (Figure \ref{fig:refcnn}).}
  \label{tab:iou}
\end{table}

\section*{Discussion}

The results of the focusing analysis shown in the previous section indicate that the 
CNN can extract
focusing information from the faults within the 
image when there lacks sufficient information along the aperture angles. We observed
that we could achieve nearly the same results of a full aperture focusing analysis, 
with a very limited aperture. For the spatial locations where the results differ, 
we believe that this difference could be attributed to a few different reasons.
One reason could be due to the lack of focusing information within the physical axes of the image.
For example, comparing the different faults in Figure \ref{fig:refsegcomp}, we observe
that although the velocity is incorrect, 
the left-most fault has been largely unaffected due to the velocity error, while the central
fault is very much unfocused. The fault segmentation of the different images readily shows the
difference in focusing of the faults. In scenarios such as these, 
it will be more challenging for the CNN 
to perform an accurate focusing analysis. 
Another possible source of uncertainty that must be considered when
training an image-focusing CNN is the labeling of field data examples. While manual/interpretative 
focusing analysis has shown to be quite successful for migration scanning methods,
manual interpretation and labeling will most likely contain errors in labeling images as 
focused or unfocused. While this is true for all
supervised learning techniques, our approach is more robust to these errors 
in that at relatively little cost, we can provide
a large amount of pre-training images to the CNN in which we have perfectly 
focused and unfocused images. Moreover, the exact classification of an image patch 
as unfocused or focused is not of the utmost importance. 
Rather, we require that the CNN provide a higher relative focusing score to patches 
that are better focused. This relaxes the accuracy requirement for the CNN for 
classification and instead allows the CNN to act as a
data-driven focusing measure of seismic images with faults. Lastly, there is always the issue with
providing enough training data for the network to generalize to many images. 
We suspect that when applying the image-focusing CNN to different
seismic images, a small amount of training image patches from each new image may need to be 
provided to the CNN for training. This will be especially important in cases in which 
acquisition parameters or the geology varies greatly between surveys.

In addition to these potential sources of uncertainty in our method, we also point
out that there is certainly room for improvement of the method. Perhaps the biggest
downside to our proposed approach is that it is entirely local and applied on a patch-by-patch
basis. The main advantage and motivation behind this choice is that by remaining local, 
we can scale to full 3D prestack images (5-dimensional images) with much greater ease.
Nevertheless, future research will possibly involve also providing the entire stacked
image to the CNN for a global-context. While the patch-by-patch approach is inherently lower 
resolution than semblance, we believe that this does not hinder our approach due to the fact that
interpretative approaches will tend to be lower resolution and better follow the
geological structure within the image. 

Comparing the computational cost of our approach with that
of the semblance-based approach, we note that the cost of our approach is 
significantly higher. This is both true for training and prediction
stages of our approach. For the semblance-based approach, which effectively
requires a stack over angles and smoothing in depth (as described by 
Equation \ref{eqn:semb}) as well as a picking procedure and the smoothing
of the picks, the total estimation of $\rho(z,x)$ completed in 0.9 min
on a single core of an Intel(R) Xeon(R) Gold 6126 2.60 GHz CPU. 
In comparison, our CNN-based approach, requires similar
amounts of memory but significantly more compute resources.
To create the synthetic pre-training data, each
synthetic model can be computed independently and therefore the computation
is embarrassingly parallel. Distributing this computation across 200 CPU
cores on 9 compute nodes (of the same hardware used for the semblance computation), 
all 8,192 synthetic patches were created in 15 minutes.
For both stages of training, we trained on a single NVIDIA V-100 GPU with 16 GB of RAM
which allowed us to store all patches in GPU memory during training. For the pre-training
stage, the total time for training was 10 minutes and for the real data the training time 
was 0.5 minutes. Finally, using the same GPU, we computed the focusing scores for all
residually migrated patches within the region of interest in 1.1 min.

One way to mitigate the significant cost increase of the CNN-based approach
is to use it in conjunction with a pure semblance-based approach.
For example, in well-illuminated areas, a semblance-based approach can be used
to provide reliable estimates of the migration scanning parameter. In areas
of poor illumination and that contain focused and unfocused geological features within
the physical axes of the seismic image, the CNN-based approach can help to 
overcome the limitations of a semblance-based approach in that area. Additionally,
our method could be used in conjunction with a diffraction imaging procedure
in which the CNN could be trained with focused/unfocused diffraction images
instead of full prestack images. While this introduces an additional step into
the workflow, it eliminates the additional offset/angle axis as diffraction-imaging
approaches often operate on stacked images, which
reduces the memory and disk read/write costs associated with performing focusing analysis
on prestack images.

\section*{Conclusion}

Convolutional neural networks provide a data-driven approach for
automating image-focusing analysis which traditionally is a highly qualitative
and interpretative process. Using focusing information present within the midpoint-depth 
axes as well as curvature information along the aperture-angle axis, 
we trained a CNN to classify local image patches as focused or unfocused. Our
limited-aperture field data example demonstrates that our image-focusing CNN is able to 
correctly estimate a spatially variant migration scanning parameter 
where a traditional semblance-based approach
fails. The comparison of our approach with the semblance-based 
full-aperture focusing analysis confirms the robustness of our CNN-based approach 
to a limited aperture.

\begin{acknowledgments}
  We would like to thank Total Energies Research and Technology US and the other 
  affiliate members of the Stanford Exploration Project 
  for the financial support of this research. We would also like to thank Rami Nammour
  for useful comments and suggestions.
\end{acknowledgments}

\bibliographystyle{seg}
\bibliography{paper}

\end{document}